\documentclass[aps,prd,eqsecnum,showpacs,amsmath,amssymb,superscriptaddress,nofootinbib,preprint]{revtex4-2}

\usepackage{graphicx}
\usepackage{dcolumn}
\usepackage{bm}
\usepackage{color}
\usepackage{subfigure}

\newcommand{\be}{\begin{equation}}
\newcommand{\ee}{\end{equation}}
\newcommand{\bea}{\begin{eqnarray}}
\newcommand{\eea}{\end{eqnarray}}

\newcommand{\bseq}{\begin{subequations}}
\newcommand{\eseq}{\end{subequations}}

\allowdisplaybreaks[1]
\begin{document}

\title{Anti-de Sitter neutron stars in the theory of gravity with nonminimal derivative coupling}

\author{Pavel E. Kashargin}
\email{pkashargin@mail.com}
\affiliation{Institute of Physics, Kazan Federal University, Kremliovskaya str. 16a, Kazan 420008, Russia}

\author{Sergey V. Sushkov}
\email{sergey$_\,$sushkov@mail.ru}
\affiliation{Institute of Physics, Kazan Federal University, Kremliovskaya str. 16a, Kazan 420008, Russia}

\date{\today}

\begin{abstract}
We consider neutron star configurations in the scalar-tensor theory of gravity with the coupling between the kinetic term of a scalar field and the Einstein tensor (such the model is a subclass of Horndeski gravity). Neutron stars in this model were studied earlier for the special case with a vanishing ``bare'' cosmological constant, $\Lambda_0=0$, and a vanishing standard kinetic term, $\alpha=0$. This special case is of interest because it admits so-called stealth configuration, i.e. vacuum configuration with nontrivial scalar field and the Schwarzschild metric. However, generally one has $\Lambda_0\not=0$ and $\alpha\not=0$ and in this case a vacuum configuration is represented as an asymptotically anti-de Sitter (AdS) black hole solution with the nontrivial scalar field. We construct neutron star configurations in this general case and show that resulting diagrams describing the relation between mass and radius of the star essentially differ from those obtained in GR or the particular model with $\alpha=\Lambda_0=0$. Instead, the mass-radius diagrams are similar to those obtained for so-called bare strange stars when a star radius decreases monotonically with decreasing mass. We show also that neutron stars in the theory of gravity with nonminimal derivative coupling are more compact comparing to those in GR or the particular model with $\alpha=\Lambda_0=0$ and suggest a way to estimate possible values of the parameter of nonminimal coupling $\ell$. {At last, using the Regge-Wheeler method, we discuss briefly the stability of obtained neutron star configurations.} 
\end{abstract}

\maketitle

\section{Introduction}
Einstein’s theory of general relativity (GR) has passed all experimental tests in its centennial history in flying colors and is very successful in describing gravitational effects in our Solar System, i.e. in the weak-field/slow-motion regime \cite{Will}. On the other hand, in recent years there appear various observational evidence and theoretical arguments which motivate strong efforts to develop modified theories of gravity which differ from GR in the infrared and ultraviolet regimes, while being consistent with observational constraints at intermediate energies \cite{Review_Berti_etal}.  
To date, many different versions of modified or extended theories of gravity have been proposed (see surveys 
\cite{Review_Salvatore:2011, Review_Clifton_etal, Review_ModGrav:2013, Review_Berti_etal, Review_Nojiri:2017, Review_Langlois:2019} and references therein). One of such models intensively studied today is Horndeski theory of gravity \cite{Horndeski} derived in the 1970s as an attempt to obtain the most general action for a scalar-tensor theory with a single scalar degree of freedom and second-order field equations. In 2011 Horndeski gravity has been rediscovered in the context of generalized Galileon theories \cite{Kobayashi:2011}, and since the interest in this model has only growing.\footnote{The literature dedicated to various aspects of Horndeski gravity is very vast, and its survey lays out of the scope of this work. The reader interesting in this topic can find some references in the already mentioned surveys \cite{Review_Berti_etal, Review_Clifton_etal}.}

Compact astrophysical objects such as black holes (BHs) and neutron stars (NSs) possess strong gravitational fields and need a relativistic theory of gravity for their adequate description. For this reason, the study of such objects could provide strong constraints -- both observational and theoretical -- on parameters of modified theories of gravity used for their modeling.
Generally, neutron stars are compact objects with a mass $M \sim 1.4 M_\odot$, a radius $R \sim 12\ {\rm km}$, and a central density as high as 5 to 10 times the nuclear equilibrium density $n_0 \approx 0.16\ {\rm fm}^{-3} $ of neutrons and protons found in laboratory nuclei ($\rho_n \approx 2.3{\rm -}2.8 \times 10^{14}\ {\rm g / cm^3}$) \cite{Book_HPY, LattimerPrakash, Schmitt}.\footnote{In this work we will use an average value for the nuclear density, $\rho_n=2.5 \times 10^{14}\ {\rm g / cm^3}$ (see \cite{Schmitt, PageReddy:2006}). }
{
	Neutron stars have been investigated in various modified theories of gravity including, in particular, 
	{
		$f(R)$ gravity 
		\cite{Babichev:2009, Babichev:2010, Cooney:2010, Orellana:2013, Ryotaro:2019, Astashenok:2021peo, Astashenok:2020qds, Astashenok:2021xpm, Astashenok:2021btj},
	}  
	$f(R,T)$ gravity \cite{Moraes:2016, Pace:2017, Mathew:2020, Carvalho2020, Pretel:2021}, 
	teleparallel gravity \cite{Ilijic:2018, Lin:2022},  
	Einstein-Dilaton-Gauss-Bonnet gravity \cite{Pani:2011, Doneva:2019}, 
	scalar-tensor gravity \cite{OdintsovOikonomou, Raissa, Horbatsch:2011, Doneva:2020},  
	massive gravity \cite{Rosca}, 
	Rastall gravity \cite{Oliveira:2015}, 
	Eddington-inspired Born-Infeld gravity \cite{HarkoLoboSushkov},
	Ho\v{r}ava–Lifshitz gravity \cite{Greenwald:2010} and etc (one can find more references, for example, in the review \cite{Olmo:2020}).
}

{
Neutron stars in Horndeski gravity have been also widely discussed in the literature 
\cite{Rinaldi:2015, Rinaldi:2016, Silva:2016, Maselli:2016, Eickhoff:2018,  Lehebel:2017fag}. Here let us note especially the important general result obtained in Ref. \cite{Lehebel:2017fag}, where was proven the absence of scalar hair for spherically symmetric and static stars in Horndeski and beyond theories involving a shift-symmetric scalar field and minimally coupled matter fields. We will see below that our stellar models are consistent with this result. 
}

%

The important subclass of Horndeski gravity is represented by models with a nonminimal derivative coupling of a scalar field with the Einstein tensor with the action
\begin{equation}\label{action1}
	S=\int d^4x\sqrt{-g}\,
	\left[\frac{1}{2\kappa} (R-2\Lambda_0)-\frac12\left( \alpha g_{\mu\nu}+\beta G_{\mu\nu} \right)\nabla^{\mu}\phi\nabla^{\nu}\phi\right]+S^{(m)},
\end{equation}
where $R$ and $G_{\mu\nu}$ are the Ricci scalar and the Einstein tensor, respectively,  $\kappa=8\pi G/c^4$ is the Einstein gravitational constant, and $S^{(m)}$ is the action for ordinary matter fields, supposed to be minimally coupled to gravity in the usual way.
Coefficients $\alpha$ and $\beta$ are real parameters, where $\alpha$ corresponds to the usual kinetic term of the scalar field, while $\beta$  determines its modified part.
$\Lambda_0$ is a `bare' (i.e. unobserved) cosmological constant. As we will see later, an observed cosmological constant $\Lambda_{AdS}$ appears as a certain combination of $\Lambda_0$ and the parameter of nonminimal derivative coupling $\beta$.  

The theory (\ref{action1}) has very interesting cosmological properties \cite{Sushkov:2009, Sushkov:2010, Sushkov:2012, Sushkov:2016, Sushkov:2020}, and provides black hole \cite{Rinaldi:2012, Minamitsuji:2013, Anabalon:2014, Babichev:2014, Kobayashi:2014, Babichev:2015} and wormhole \cite{Sushkov:2012b, Sushkov:2014} solutions. Neutron stars have been also explored within the model with nonminimal derivative coupling \cite{Rinaldi:2015, Rinaldi:2016, Silva:2016, Maselli:2016, Eickhoff:2018} 		
For the first time, Cisterna et al. \cite{Rinaldi:2015} constructed asymptotically flat neutron stars in the theory (\ref{action1}) for the so-called stealth configuration with $\Lambda_0=\alpha=0$ \cite{Babichev:2014}.
Later, Cisterna et al. extended their work to more realistic equations of state and slowly rotating solutions in \cite{Rinaldi:2016}. Independently, in 2016 Maselli et al. conducted a similar study in \cite{Maselli:2016}.
Recently, Blázquez-Salcedo and Eickhoff \cite{Eickhoff:2018} calculated the spectrum of axial quasinormal modes of static and spherically symmetric neutron stars obtained in \cite{Rinaldi:2015}.

It is worth noticing that the specific case $\Lambda_0=\alpha=0$ mentioned above means that the bare value of the cosmological constant $\Lambda_0$ is setting zero by hand, and the usual kinetic term of the scalar field, $\frac12 g_{\mu\nu}\nabla^\mu\phi\nabla^\nu\phi$, is assumed to be absent. 
Namely this case leads to vacuum configurations with a non-trivial scalar field and the Schwarzschild metric. Such solutions were dubbed stealth configurations in \cite{Babichev:2014}. However, generally in vacuum the theory described by the action (\ref{action1}) leads to asymptotically anti-de Sitter (AdS) black hole solutions with a nontrivial scalar field configuration \cite{Rinaldi:2012, Minamitsuji:2013, Anabalon:2014, Babichev:2014, Kobayashi:2014, Babichev:2015}. 

Our goal is to explore neutron star configurations with AdS asymptotic within the framework of the full theory (\ref{action1}) without imposing any restrictions on the parameters $\Lambda_0$ and $\alpha$.

The paper is organized as follows. In Section \ref{sec2} we derive general equations describing an external and internal configuration of a neutron star in the theory of gravity with nonminimal derivative coupling. The external vacuum solution is analyzed in Section \ref{sec3}. A detail analysis of interior of the star and constructing of a complete solution joining internal and external configurations is provided in Section \ref{sec4}. {In Section \ref{sec5} we discuss briefly the stability of neutron star configurations.} Obtained results are summarized in Section \ref{Summary}.

\section{Basic equations} \label{sec2}

\subsection{Action and field equations}
Let us consider a gravitational theory of a real scalar field $\phi$ with nonminimal derivative coupling to the curvature which is described by the action (\ref{action1}).
Note that we can get rid of the parameter $\alpha$ by redefining the scalar field as follows $|\alpha|^{1/2}\phi\to \phi$. In this case the kinetic term $\frac12\alpha g_{\mu\nu}\nabla^{\mu}\phi\nabla^{\nu}\phi$ takes the form $\frac12\varepsilon_1 g_{\mu\nu}\nabla^{\mu}\phi\nabla^{\nu}\phi$, where $\varepsilon_1$ is the sign of $\alpha$, i.e. $\varepsilon_1=\pm 1$. The parameter $\beta$ has the dimension $(length)^2$, and so it will be convenient to use the notation $\beta=\varepsilon_2 \ell^2$, where $\varepsilon_2$ is the sign of $\beta$, i.e. $\varepsilon_2=\pm 1$, and $\ell$ is a characteristic length which characterizes the nonminimal derivative coupling between the scalar field and curvature. 
Ultimately, we can rewrite the action (\ref{action1}) as follows
\begin{equation}\label{action}
S=\int d^4x\sqrt{-g}\,
\left[\frac{1}{2\kappa} (R-2\Lambda_0)-\frac12\left( \varepsilon_1 g_{\mu\nu}+\varepsilon_2 \ell^2 G_{\mu\nu} \right)\nabla^{\mu}\phi\nabla^{\nu}\phi\right]+S^{(m)}.
\end{equation}
Now, varying the action (\ref{action}) with respect to the metric $g_{\mu\nu}$, we obtain the following gravitational field equations:
\be\label{gfe}
\frac1\kappa\, (G_{\mu\nu}+g_{\mu\nu}\Lambda_0)=\varepsilon_1 T_{\mu\nu}^{(\phi)}+\varepsilon_2\ell^2 \Theta_{\mu\nu}+T_{\mu\nu}^{(m)},
\ee
where
\begin{eqnarray}
T_{\mu\nu}^{(\phi)}&=&\nabla_{\mu}\phi\nabla_{\nu}\phi-\dfrac{1}{2}g_{\mu\nu}\left(\nabla\phi\right)^2,
\\
\Theta_{\mu\nu}&=&-\frac{1}{2}\nabla_{\mu}\phi\nabla_{\nu}\phi R+2\nabla_{\alpha}\phi\nabla_{(\mu}\phi R^{\alpha}_{\nu)}+\nabla^{\alpha}\phi\nabla^{\beta}\phi R_{\mu\alpha\nu\beta} +
\nonumber\\
&&\nabla_{\mu}\nabla^{\alpha}\phi\nabla_{\nu}\nabla_{\alpha}\phi-\nabla_{\mu}\nabla_{\nu}\phi\Box\phi-\frac{1}{2}\left(\nabla\phi\right)^2G_{\mu\nu}+\nonumber\\
&&g_{\mu\nu}\left[-\frac{1}{2}\nabla^{\alpha}\nabla^{\beta}\phi\nabla_{\alpha}\nabla_{\beta}\phi+\frac{1}{2}\left(\Box\phi\right)^2-\nabla_{\alpha}\phi \nabla_{\beta}\phi R^{\alpha\beta}\right],
\end{eqnarray}
and $T_{\mu\nu}^{(m)}$ is a stress-energy tensor of the ordinary matter. Hereafter as the matter we will consider a perfect fluid with the energy-momentum tensor 
\be 
T_{\mu\nu}^{(m)}=(\epsilon + p)u_{\mu}u_{\nu}+p g_{\mu\nu},
\ee
where $u_{\mu}$ is a unit timelike 4-vector, $u_\mu u^\mu=-1$, $\epsilon$ is an energy density, and $p$ is an isotropic pressure. The equation of motion of the perfect fluid is given by the conservation law:
\be\label{mfe}
\nabla^{\mu} T_{\mu\nu}^{(m)}=0.
\ee
As well, varying the action (\ref{action}) with respect to $\phi$, we obtain the equation of motion of the scalar field:
\be\label{sfe}
\nabla_{\mu} J^{\mu}=0,
\ee
where
\begin{equation}
J^{\mu}=\left(\varepsilon_1 g^{\mu\nu}
+\varepsilon_2\ell^2 G^{\mu\nu}\right)\nabla_{\nu}\phi.
\end{equation}

\subsection{Field equations for static spherically symmetric configurations}
In this section we will focus on static spherically symmetric configurations in the theory (\ref{action}). A general static spherically symmetric spacetime metric can be represented in the following form:
\be
ds^2=-A(r)c^2dt^2+\frac{dr^2}{B(r)}+r^2\left(d\theta^2+\sin^2\theta d\varphi^2\right),
\ee
where $A(r)$ and $B(r)$ are functions of the radial coordinate $r$. Assume also that the scalar field $\phi$, the energy density $\epsilon$, and the pressure $p$ depend only on $r$, i.e. $\phi=\phi(r)$, $\epsilon=\epsilon(r)$, and $p=p(r)$.

Now nonzero independent components of the gravitational field equations (\ref{gfe}) take the following form:
\begin{eqnarray}
\label{fe1}
\frac1\kappa \left(-\frac{B'}{r}+\frac{1-B}{r^2}\right)
&=&
\epsilon +\frac1\kappa\Lambda_0 +\frac12\varepsilon_1 B \psi^2
-\varepsilon_2\ell^2\frac{B\psi^2}{2r^2}\left( 1+B+3rB'+4rB\frac{\psi'}{\psi} \right),
\\
\label{fe2}
\frac1\kappa \left(\frac{B A'}{r A}-\frac{1-B}{r^2}\right)
&=&
p-\frac1\kappa\Lambda_0 +\frac12\varepsilon_1 B{\psi^2}
-\varepsilon_2\ell^2\frac{B\psi^2}{2r^2}\left(1-3B-3rB\frac{A'}{A}\right),
\end{eqnarray}
where a prime means a derivative with the respect of $r$, and $\psi=\phi'$. 

The equation of motion of the perfect fluid (\ref{mfe}) yields
\be\label{eom}
\frac{A'}{A}=-\frac{2p'}{\epsilon +p}.
\ee

{
The scalar field equation (\ref{sfe}) can be integrated, and the first integral reads
\begin{equation}\label{seom}
\left[ \varepsilon_1 r^2 -\varepsilon_2\ell^2\left(1-B-rB\frac{A'}{A}\right) \right]\psi \sqrt{AB}=Q,
\end{equation}
where $Q$ is a constant of integration, which plays a role of scalar charge, in the sense that it determines the far-away behavior of the scalar field.
}

\subsection{Equation of state}
The system of four field equations (\ref{fe1})-(\ref{eom}) contains five unknown functions $A(r)$, $B(r)$, $\psi(r)$, $p(r)$, and $\epsilon(r)$, and hence is not completed. To make it complete, one needs to add an equation of state relating the pressure and the energy density.
In this paper, we will consider the polytropic equation of state\footnote{
	The polytropic equation of state (EoS) is one of the most venerable EoS used in the context of Newtonian and relativistic theory to deal with
	a variety of astrophysical scenarios (see \cite{ShapiroTeukolsky} and references therein). In our knowledge, the EoS in the form (\ref{eos}) for the neutron stars description was firstly considered by Tooper \cite{Tooper:1965} in 1965.}
\be\label{eos}
p=K \rho_0^{\Gamma}, 
\quad \epsilon = \rho_0 c^2 + \frac{p}{\Gamma-1},
\ee 
where $\rho_0$ is a baryonic mass density, 
$\Gamma = 1 + 1/n$ is the adiabatic index, $n$ is the polytropic index, and $K$ is the polytropic constant encoding the temperature and the entropy per nucleon, as well as the star chemical composition.
Note that excluding $\rho_0$ from (\ref{eos}) yields
\be\label{eos_bab}
\epsilon=c^2 \left(\frac{p}{K}\right)^{1/\Gamma}+\frac{p}{\Gamma-1}.
\ee
Note that in this work we use $\Gamma=2$ and $K=1.79\times 10^5\, {\rm cgs}$ since these values lead to compact objects with accepted mass and radius of neutron stars \cite{Rinaldi:2015}.

\subsection{Boundary conditions}
The equations (\ref{fe1})--(\ref{eos}) form a closed system of five ordinary differential equations for five function $A(r)$, $B(r)$, $\phi(r)$, $p(r)$, and $\epsilon(r)$. Boundary conditions for the system (\ref{fe1})--(\ref{eos}) are usually determined at the center of a star, $r=0$. Expanding the functions $A(r)$, $B(r)$, $\phi(r)$, $p(r)$, $\epsilon(r)$ around $r=0$ and substituting the expansions into Eqs. (\ref{fe1})--(\ref{eos}), we can find that the regularity conditions at the center of star dictate that $Q=0$. 
{
	Here it is worth stressing that the condition $Q=0$ means that the scalar charge is equal to zero, and hence a possible stellar configuration has no scalar hairs in a total consistence with the result proved in \cite{Lehebel:2017fag}.
} 
The other conditions read
\be
A_c'=0, \quad B_c=1,\quad B_c'=0, \quad \psi_c=0, \quad p_c'=0,
\ee
where the lowercase index `c' marks a value at the center, i.e. $A_c=A(0)$, etc. 
Note that the value of $A_c$ will be fixed after matching internal and external solutions at the star boundary.
Therefore, the only free parameter is the value of the pressure in the center of the star, $p_c$. In turn, by using the EoS, the central pressure $p_c$ can be expressed in term of the central baryonic mass density $\rho_{0c}$ as follows $p_c=K\rho_{0c}^\Gamma$.

\subsection{Normal system of field equations}
For the purpose of numerical analysis it will be convenient to represent the system (\ref{fe1})--(\ref{eos}) in a normal dimensionless form. Introducing the dimensionless values as follows
\be\label{dlvalues}
\xi=\Lambda_0\ell^2, \quad
x=\frac{r}{\ell},\quad 
{\cal E}=\kappa\ell^2 \epsilon, \quad
{\cal P}=\kappa\ell^2 p, \quad
\Psi^2=\kappa\ell^2 \psi^2,
\ee
and resolving the system (\ref{fe1})--(\ref{eos}) with respect to highest derivatives, one can obtain
\begin{eqnarray}
\nonumber
&& \frac{dB}{dx} = -\frac{1}{\Delta}
\bigg[\Big((1+\varepsilon \xi)x^4 +(\varepsilon-5\xi) x^2 +2\Big)B +\Big((1-\varepsilon x^2) {\cal E}
+2 (3-\varepsilon x^2) {\cal P} \Big) x^2B
\\ 
&&~~~~~~~~~~~~ 
-(1-\varepsilon x^2)^2 \,
\big(2 -(\varepsilon+\xi) x^2 -x^2 {\cal E}  \big)\bigg]
\\
&& \frac{d{\cal P}}{dx} =-\frac{({\cal E} + {\cal P}) (1-B-\varepsilon x^2) }{2xB},
\\
&& \frac{dA}{dx}=\frac {A(1-B-\varepsilon x^2)}{xB},
\\
&& \Psi^2=- \frac{ x^2 (\varepsilon-\xi +{\cal P})} {\varepsilon_2 B\, (1-\varepsilon x^2)},
\end{eqnarray}
where $\varepsilon=\varepsilon_1/\varepsilon_2$ and
$$
\Delta =x(1-\varepsilon x^2 ) \big(2 -(\varepsilon +\xi) x^2 +x^2 {\cal P}\big).
$$

Note that in order to provide a regularity of solutions of (\ref{eqB_dl})--(\ref{eqPsi_dl}) on the entire interval of varying the radial coordinate, $x\in[0,\infty)$, we need to make a choice $\varepsilon=\varepsilon_1/\varepsilon_2=-1$, because in this case $(1-\varepsilon x^2)=(1+x^2)>0$, and denominators in  (\ref{eqB_dl}) and (\ref{eqPsi_dl}) do not go to zero. The choice $\varepsilon=-1$ means that $\varepsilon_1$ and $\varepsilon_2$ have different signs and, ultimately, it means that the usual kinetic term $\alpha g_{\mu\nu} \nabla^\mu\phi\nabla^\nu\phi$ and the modified term $\beta G_{\mu\nu} \nabla^\mu\phi\nabla^\nu\phi$ enter into the Lagrangian (\ref{action1}) with different signs.

Now, assuming $\varepsilon=-1$, we find
\begin{eqnarray}
	\nonumber\label{eqB_dl}
	&& \frac{dB}{dx} = -\frac{1}{\Delta}
	\bigg[\Big((1- \xi)x^4 -(1+5\xi) x^2 +2\Big)B +\Big((1+ x^2) {\cal E}
	+2 (3+ x^2) {\cal P} \Big) x^2B
	\\ 
	&&~~~~~~~~~~~~ 
	-(1+ x^2)^2 \,
	\big(2 +(1-\xi) x^2 -x^2 {\cal E}  \big)\bigg]
	\\\label{eqP_dl}
	&& \frac{d{\cal P}}{dx} =-\frac{({\cal E} + {\cal P}) (1-B+x^2) }{2xB},
	\\\label{eqA_dl}
	&& \frac{dA}{dx}=\frac {A(1-B+x^2)}{xB},
	\\\label{eqPsi_dl}
	&& \Psi^2= \frac{ x^2 (1+\xi- {\cal P})} {\varepsilon_2 B\, (1+x^2)},
\end{eqnarray}
where 
$$
\Delta =x(1+x^2 ) \big(2 +(1-\xi) x^2 +x^2 {\cal P}\big).
$$
Note also that the sign $\varepsilon_2=\pm1$ in (\ref{eqPsi_dl}) is still undefined. To be determined, $\varepsilon_2$ should provide the positivity of $\Psi^2$, that is $\varepsilon_2(1+\xi-{\cal P}) \ge 0$. 

As well, let us write down the equation of state (\ref{eos_bab}) in the dimensionless form:
\be\label{eos_dl}
{\cal E}=\kappa\ell^2 c^2 \left(\frac{\cal P}{\kappa\ell^2 K}\right)^{1/\Gamma}+\frac{\cal P}{\Gamma-1}.
\ee

\section{External vacuum solution} \label{sec3}
Outside the star one has a scalar ``vacuum'' with a nontrivial configuration of the scalar field. In this section we will consider the vacuum solution of Eqs. (\ref{eqB_dl})-(\ref{eqPsi_dl}). 
To describe this configuration we have to substitute $\rho=0$ and $p=0$, i.e. ${\cal E}=0$ and ${\cal P}=0$, into Eqs. (\ref{eqB_dl})-(\ref{eqPsi_dl}). 
As the result, we obtain  
\begin{eqnarray}
\label{vacB1}
\frac{dB}{dx} &=& -\frac{(1-\xi)x^4 -(1+5\xi)x^2+2}{x(1+ x^2) \big((1-\xi)x^2+2\big)}\, B 
+x +\frac1x.
\\\label{vacA1}
\frac{dA}{dx} &=& \frac {A(1-B +x^2)}{xB},
\\ \label{vacPsi1}
\Psi^2 &=& \frac{x^2(1+\xi)} {\varepsilon_2  B\, (1 +x^2)}.
\end{eqnarray}

In the case $\xi=-1$ the solution of these equations has the particularly simple form:
\begin{eqnarray}
A(r) = B(r) = 1-\frac{r_g}{r}+\frac{|\Lambda_{AdS}|}{3}\, r^2,
\quad \Psi^2(r)=0,
\label{stealthSS}
\end{eqnarray}
where $\Lambda_{AdS}=-1/\ell^{2}$ and the integration constants fixed as $C_1=-4 r_g/\ell=-8MG/c^2\ell$ and $C_2=1/12$. The solution (\ref{stealthSS}) represents a Schwarzschild Anti-de Sitter (SAdS) black hole with the Schwarzschild mass $M$ and the effective negative cosmological constant $\Lambda_{AdS}$.

In the general case $\xi\not=-1$ one has to choose a proper sign $\varepsilon_2=\pm1$ in order to provide positivity of $\Psi^2$. From Eq. (\ref{vacPsi1}) we obtain $\varepsilon_2=+1$ if $\xi>-1$, while $\varepsilon_2=-1$ if $\xi<-1$. Then, solving Eqs. (\ref{vacB1}) and (\ref{vacA1}) yields 
\begin{eqnarray}
	\label{vacsol_B}
	&& B(x) =\frac{(x^2+1)^2}{\big((1-\xi)x^2+2\big)^2}\, F(x),
	\\ 
	&& A(x)=3 C_2\, F(x),
	\label{vacsol_A}
\end{eqnarray}
where 
$$
F(x)=(1-\xi)(3+\xi)+\frac{1}{x}\left((1+\xi)^2 \arctan x+C_1\right) +\frac{x^2}{3}(1-\xi)^2,
$$
and $C_1$ and $C_2$ are constants of integration.  
Since we are interesting in an external solution, we have to consider asymptotics of $B(x)$, $A(x)$ and $\Psi^2(x)$ at $x\to\infty$. 
As the result, we have
\begin{eqnarray}
\label{asB}
B(x) &=& \frac{x^2}{3}+\frac{7+\xi}{3(1-\xi)} +\frac{C_1+\frac12(1+\xi)^2\pi}{(1-\xi)^2}\,\frac1x
+{\cal O}(x^{-2}),
\\
\label{asA}
A(x) &=& 3C_2(1-\xi)(3+\xi)\left[1+\frac{1-\xi}{3(3+\xi)}x^2 +\frac{C_1+\frac{\pi}{2}(1+\xi)^2}{(3+\xi)(1-\xi)}\,\frac1x\right]
+{\cal O}(x^{-2}),\\
\Psi^2(x) &=& \frac{3(1+\xi)}{\varepsilon_2 x^2} +{\cal O}(x^{-4}).
\end{eqnarray}
The asymptotical form of $B(x)$ and $A(x)$ given by Eqs. (\ref{asB}) and (\ref{asA}) shows that one has anti-de Sitter-Schwarzschild spacetime geometry outside the star. 
Assuming that $t$ is the time of a distant observer, one can fix the value of $C_2$ as follows $3C_2(1-\xi)(3+\xi)=1$. Additionally, one has to demand $C_2>0$ in order to guarantee the same sign of the metric functions $B(x)$ and $A(x)$, hence one has $(1-\xi)(3+\xi)>0$, or 
\begin{equation}\label{range_xi}
	-3<\xi<1\,.
\end{equation}   
Returning to the dimensional radial coordinate $r=\ell x$ and choosing appropriately the constants of integration $C_1$ and $C_2$, we find that far from the star
\be\label{A_AdS}
A(r)\approx 1-\frac{r_g}{r} +\frac{|\Lambda_{AdS}|}{3}\,r^2,
\ee 
where
\be\label{C1}
r_g=\frac{2GM}{c^2}=-\ell\,\frac{C_1+\frac{\pi}{2}(1+\xi)^2}{(3+\xi)(1-\xi)},
\ee 
and
\begin{equation}
	\label{Leff}
	\Lambda_{AdS}=-\frac{1-\xi}{3+\xi}\, \frac{1}{\ell^2},
\end{equation}
is the effective negative cosmological constant

{
It is important to point out that, analyzing the solutions (\ref{stealthSS}) and (\ref{A_AdS}), we can see that the Schwarzschild solution with $A(r)=B(r)=1-r_g/r$ is recovered in the limit $\ell\to\infty$. As was first stated in Ref. \cite{Rinaldi:2012} (see also Refs. \cite{Rinaldi:2016,Minamitsuji:2013}), this fact means that $\ell$ is a nonperturbative parameter of the modified theory of gravity (\ref{action}), that is general relativity is not recovered in the limit $\ell\to 0$. Instead, the deviation from general relativity vanishes when $\ell$ diverges, i.e. $\ell\to\infty$.\footnote{Here it is worth noticing the recent work \cite{NashedSaridakis:2020}, where the authors found new static and rotating charged spherically symmetric black hole solutions in the framework of $f(\cal{R})$ gravity and shown that the obtained solutions belong to two branches, one that contains the Kerr-Newman solution of general relativity as a particular limit and one that arises purely from the gravitational modification and cannot be obtained by a perturbative way.}
Note also that earlier the nonperturbative character of $\ell$ had been found in cosmological applications \cite{Sushkov:2009, Sushkov:2010}. Namely, it had been shown that cosmological models with the nonminimal derivative coupling demonstrate at early time an universal asymptotic quasi-De Sitter behavior with the Hubble parameter $H=1/3\ell$. 
	
Assuming that the gravitational field of a spherically symmetric mass has the SAdS asymptotic (\ref{stealthSS}) or (\ref{A_AdS}), one must be sure that this does not contradict all known astronomical data. In particular, the admissible magnitude of an additional term $\Lambda_{AdS}$ must be constrained by various observable Solar system effects such as gravitational redshift, light deflection, gravitational time delay, geodetic precession, etc. Solar system effects caused by $\Lambda$-term analyzed in details in Refs. \cite{Kagramanova:2006, Jetzer:2006, Sereno:2006, Iorio:2008, IorioSaridakis:2012, Arakida:2013, XieDeng:2013, Iorio:2015}, see also \cite{Iorio:2018} and references therein. Different Solar system effects give different estimates for $\Lambda$. For example, checking extra-precession of the inner planets of the Solar system gives $|\Lambda|\lesssim 10^{-36}\text{---}10^{-37}\ {km}^{-2}$. Using this constraint and the relation (\ref{Leff}) for $\Lambda_{AdS}$, we can assume the following estimation: 
\begin{equation}
	\frac{1-\xi}{3+\xi}\, \frac{1}{\ell^2}\lesssim 10^{-37}\ {km}^{-2},
\end{equation}	
In principle, one can fulfill this constraint for {\em any} magnitudes of the characteristic length $\ell$ by choosing the free dimensionless parameter $\xi$ as close to unity as necessary. In case $O(1-\xi)\sim 1$, one has $\ell\gtrsim 10^{18}\ km\sim 30\ kpc\sim$ diameter of the Milky Way. Note also that cosmological models with nonminimal derivative coupling give rather wide range for the magnitude of $\ell$. For example, in Ref. \cite{Rinaldi:2016} it was shown that  the parameter $\ell$ can be arbitrarily large.
}

\section{Internal solution} \label{sec4}
An interior configuration of the star is described by the field equations (\ref{eqB_dl})--(\ref{eqPsi_dl}) together with the equation of state (\ref{eos_dl}). From Eq. (\ref{eqPsi_dl}) one can see that positivity of $\Psi^2$ inside the star is fulfilled provided $\varepsilon_2(1+\xi-{\cal P}) > 0$. Analyzing vacuum configurations, we found that there are two different possibilities: 
\begin{eqnarray}
	&& \textrm{(i)}\quad -3<\xi\le-1,\quad \varepsilon_2=-1;
	\nonumber\\
	&& \textrm{(ii)}\quad -1<\xi<1,\quad \varepsilon_2=+1.
	\nonumber
\end{eqnarray}
In the first case the value of $\varepsilon_2(1+\xi-{\cal P})$ is always positive. In the second case the value of $\varepsilon_2(1+\xi-{\cal P})$ is positive provided 
\begin{equation}\label{cond4P}
	{\cal P}<1+\xi. 
\end{equation}
Since the pressure is maximal at the center of the star, the last condition can be recast as follows
\begin{equation}
	{\cal P}_c=\kappa\ell^2 p_c<1+\xi,
\end{equation}
where $p_c$ is the central pressure.

\subsection{Scheme of numerical integration}
We explore internal configurations of neutron stars for different sets of model parameters $\xi$ and $\ell$ using the following scheme of numerical integration of the system (\ref{eqB_dl})--(\ref{eos_dl}). 
First, we find solutions for $B(x)$ and ${\cal P}(x)$ integrating Eqs (\ref{eqB_dl}) and (\ref{eqP_dl}) from the center $r=0$ to the boundary of the star $r=R$ with the following initial conditions: $B(0)=B_c=1$, ${\cal P}(0)={\cal P}_c=\kappa\ell^2 K\rho_{0c}^\Gamma$. The boundary of the star is defined as ${\cal P}(R)=0$. At the boundary the internal solution $B_{in}(x)$ is matching with the external vacuum solution $B_{vac}(x)$ given by (\ref{vacsol_B}), i.e. $B_{in}(R)=B_{vac}(R)$. From the matching condition we fix the constant of integration $C_1$ and, ultimately, the Schwarzschild mass $M$ given by Eq. (\ref{C1}). Then, with solutions found for $B(x)$ and ${\cal P}(x)$, we obtain $A(x)$ from (\ref{eqA_dl}), and $\Psi^2(x)$ from (\ref{eqPsi_dl}). A constant of integration of Eq. (\ref{eqA_dl}) is fixed by the matching condition at the boundary of star, $A_{in}(R)=A_{vac}(R)$, where the vacuum solution $A_{vac}(x)$ is given by (\ref{vacsol_A}).

\subsection{Results of numerical integration}

\subsubsection{The case $\xi=-1$ }

First let us consider in detail the case $\xi=-1$. This case corresponds to the special choice of model parameters of the theory (\ref{action}), such that $\Lambda_0=-\ell^{-2}$, and in this case the vacuum solution has the particularly simple form (\ref{stealthSS}) corresponding to the Schwarzschild-anti de Sitter black hole. Also, it follows from Eq. (\ref{eqPsi_dl}) that $\varepsilon_2=-1$ and, since $\varepsilon=\varepsilon_1\varepsilon_2=-1$, one has $\varepsilon_1=+1$. This choice of signs means that we consider the theory (\ref{action1}) with an ordinary positive kinetic term, $\alpha>0$, and a negative nonminimal derivative coupling, $\beta<0$.   

\begin{figure}[ht]
	\begin{center}
		\includegraphics[scale=0.4]{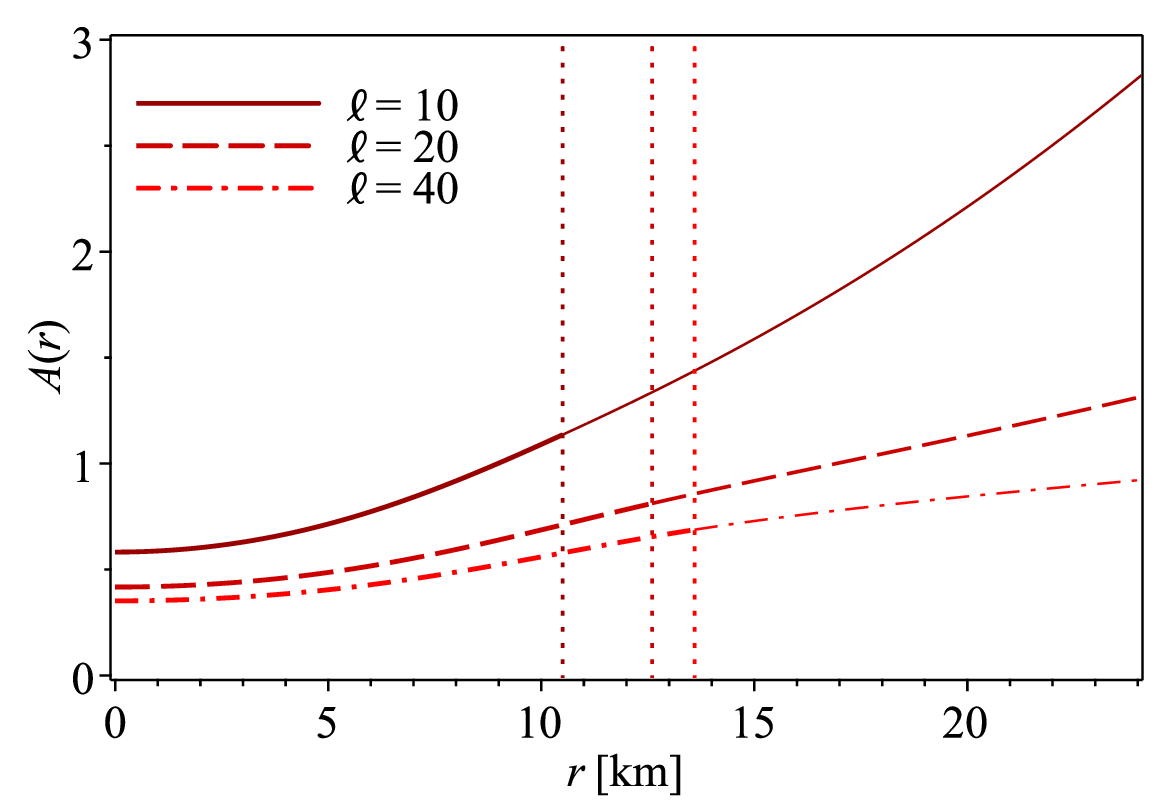}
		\includegraphics[scale=0.4]{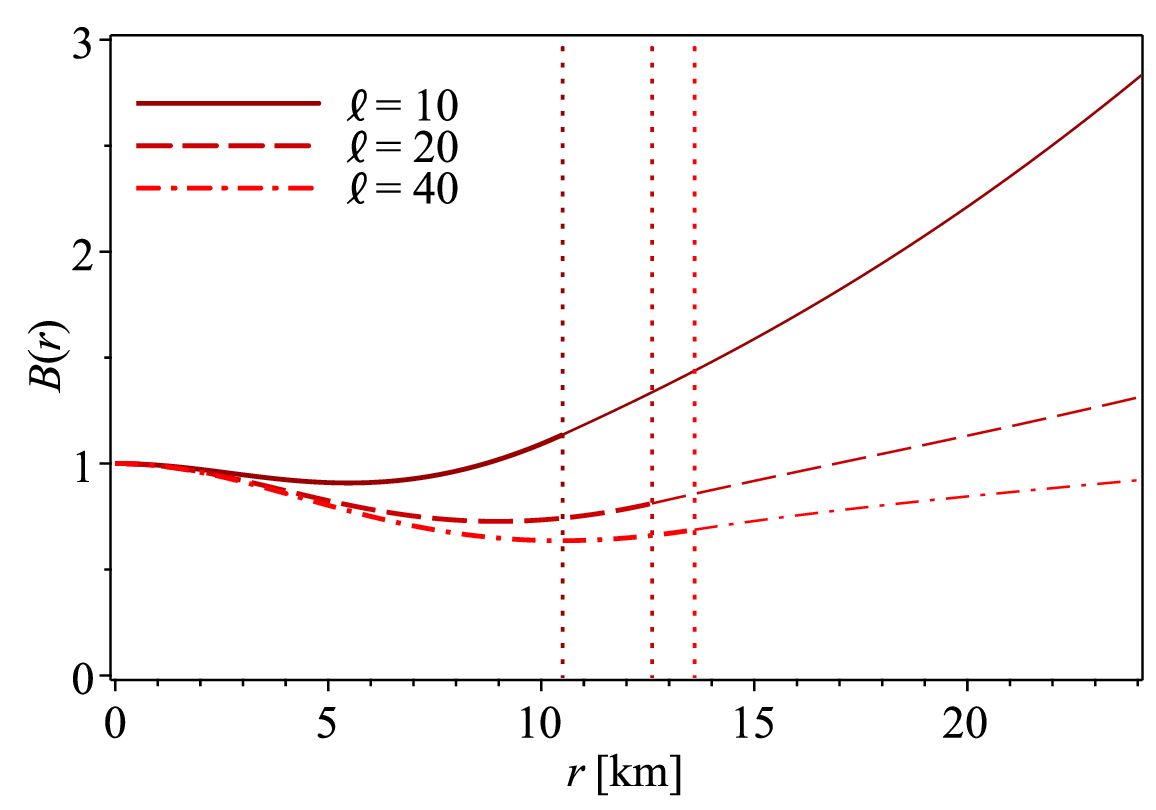}\\
		\includegraphics[scale=0.4]{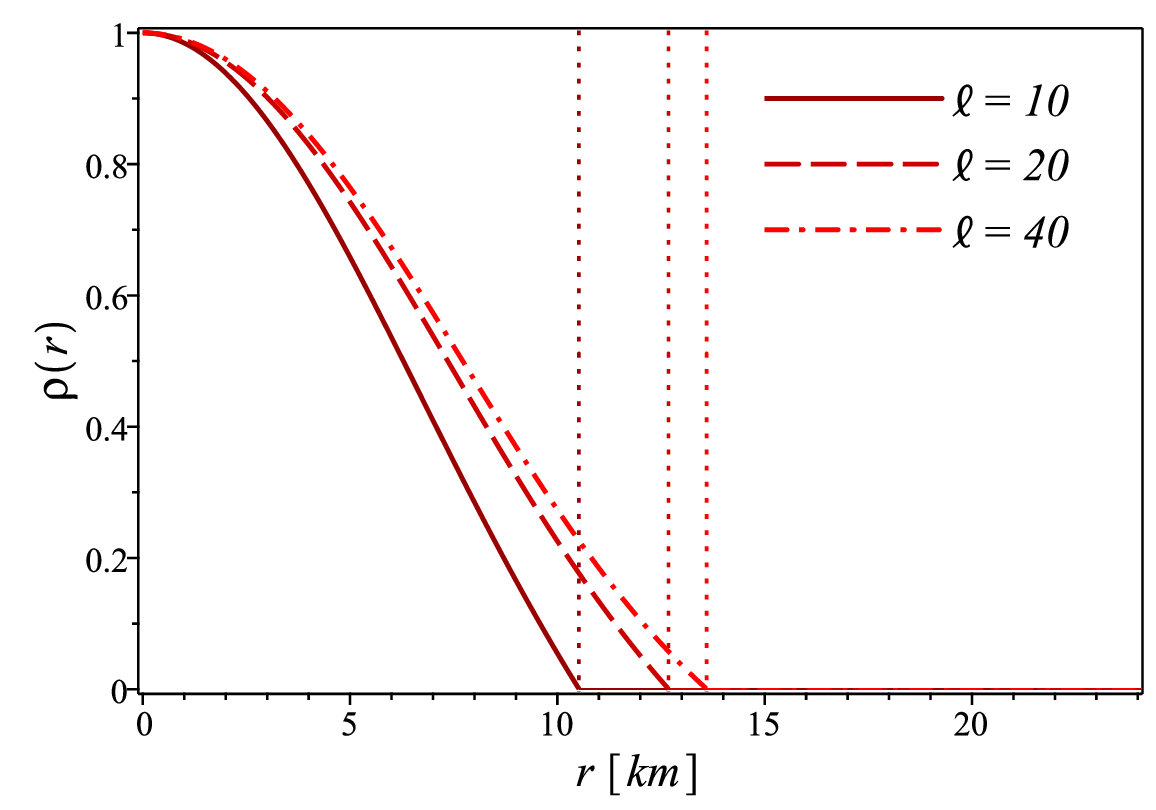}
		\includegraphics[scale=0.4]{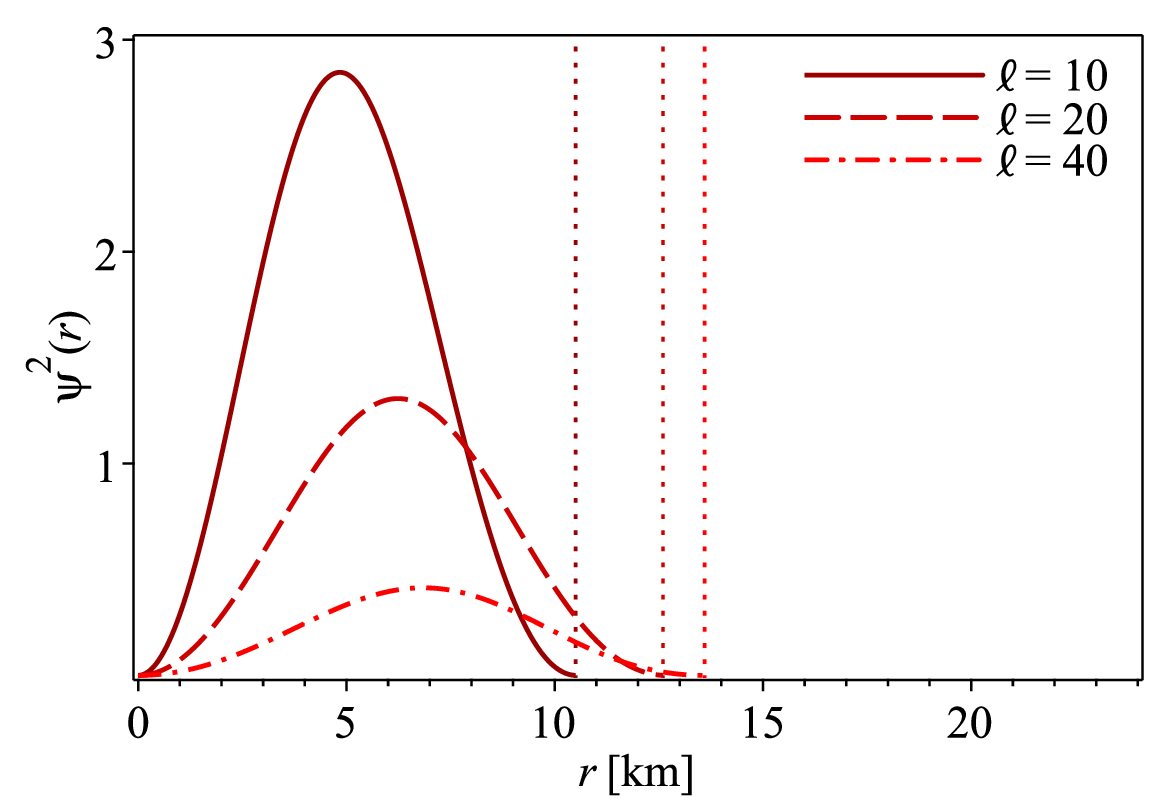}
	\end{center}	
	\caption{Graphs of the functions  $A(r)$, $B(r)$, $\rho(r)$ and $\Psi^2(r)$ in the case $\xi=-1$ are shown for three values of the nonminimal derivative coupling parameter $\ell=10, 20, 40\ km$ and the central baryonic mass density $\rho_{0c}=1.0\times 10^{15}\ g/cm^3$. Vertical dotted lines mark the boundary of star ($\ell=10, 20, 40$ from left to right).  
	\label{ABRhoPsi}}
\end{figure}

\begin{figure}[th]
	\begin{center}
		\includegraphics[scale=0.41]{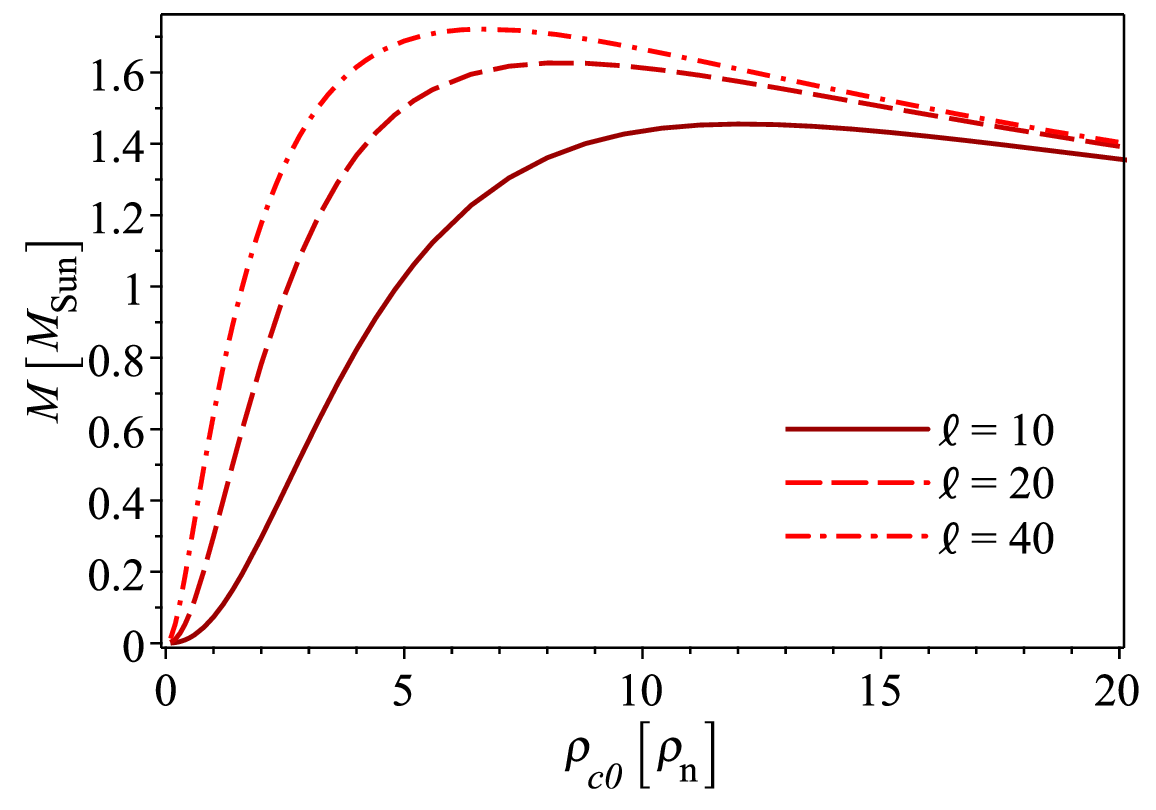}\ 
		\includegraphics[scale=0.41]{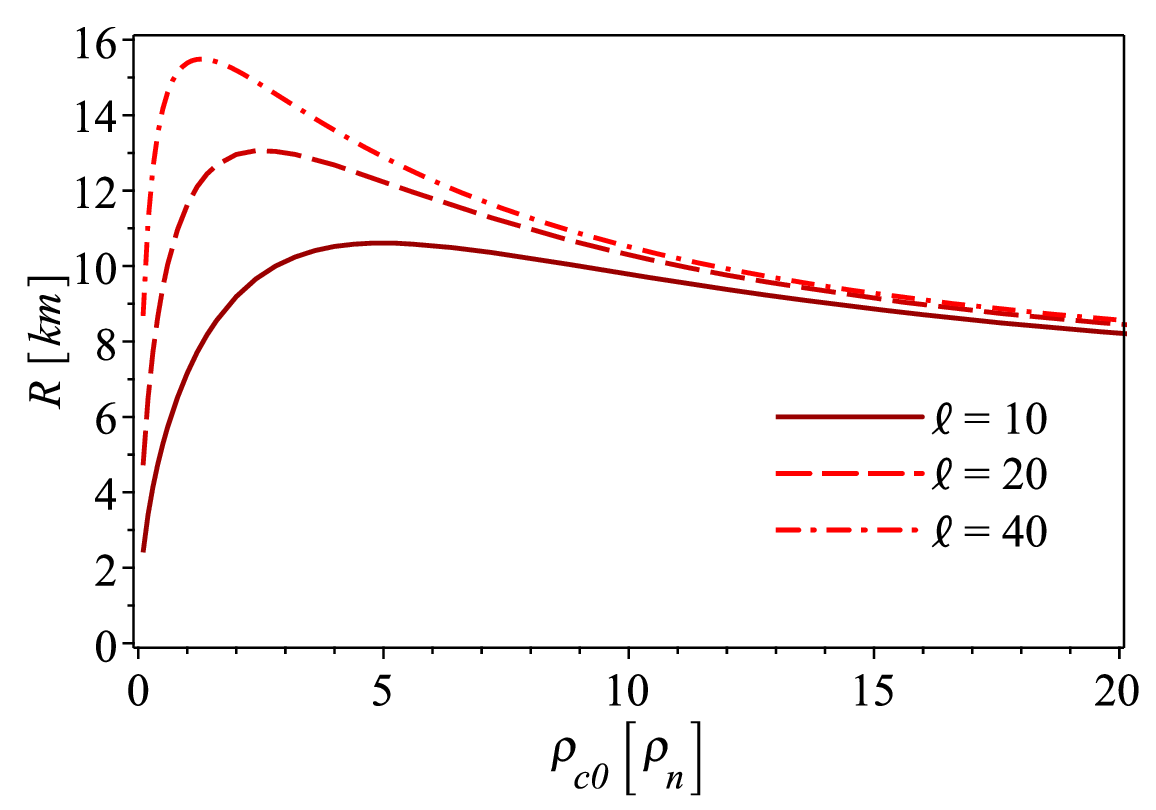}
	\end{center}	
	\caption{\label{MR_rho_xi-1}
		The dependence of the mass $M$ (left panel) and the radius $R$ (right panel) on the central baryonic mass density $\rho_{0c}$ in the case $\xi=-1$ is shown for three values of the nonminimal derivative coupling parameter $\ell=10, 20, 40\ {\rm km}$. Values of $\rho_{0c}$ are given in term of the nuclear density $\rho_n=2.5\times10^{14}\ {\rm g/cm}^3$.} 
\end{figure}

\begin{figure}[th]
\begin{center}
	\includegraphics[scale=0.4]{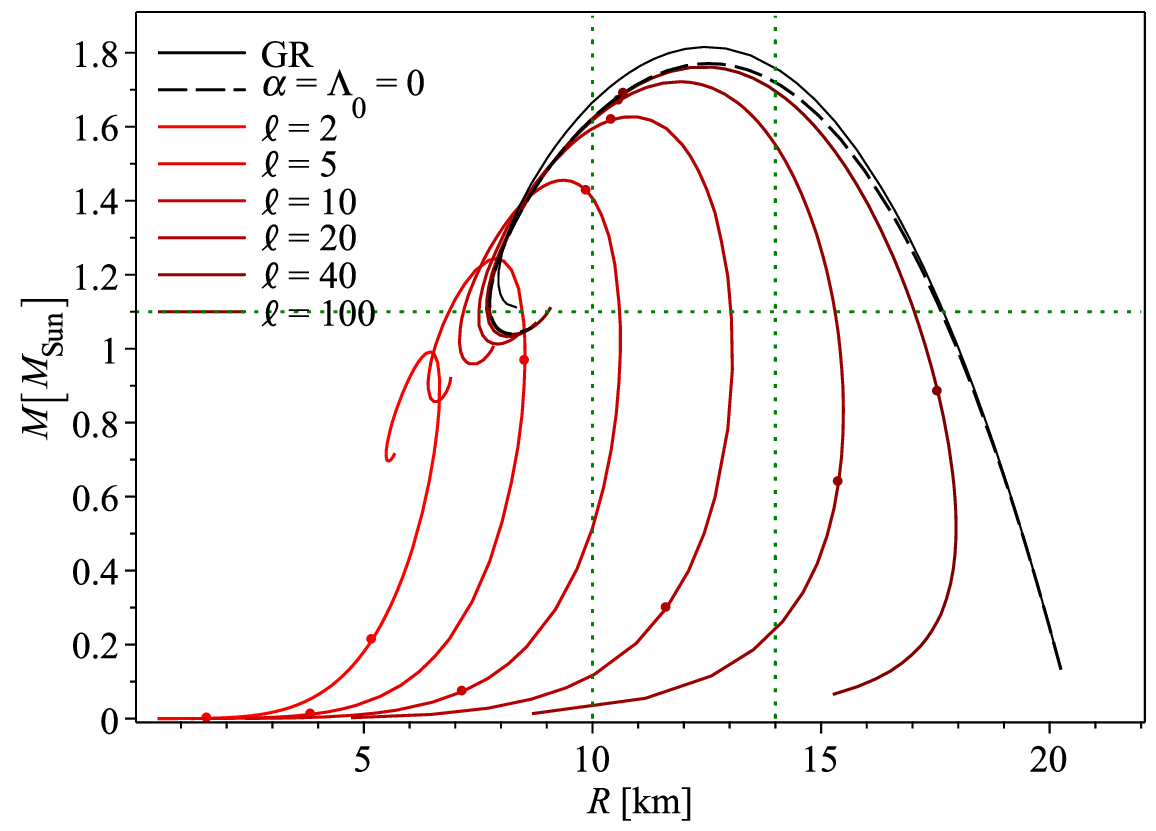}\ 
	\includegraphics[scale=0.4]{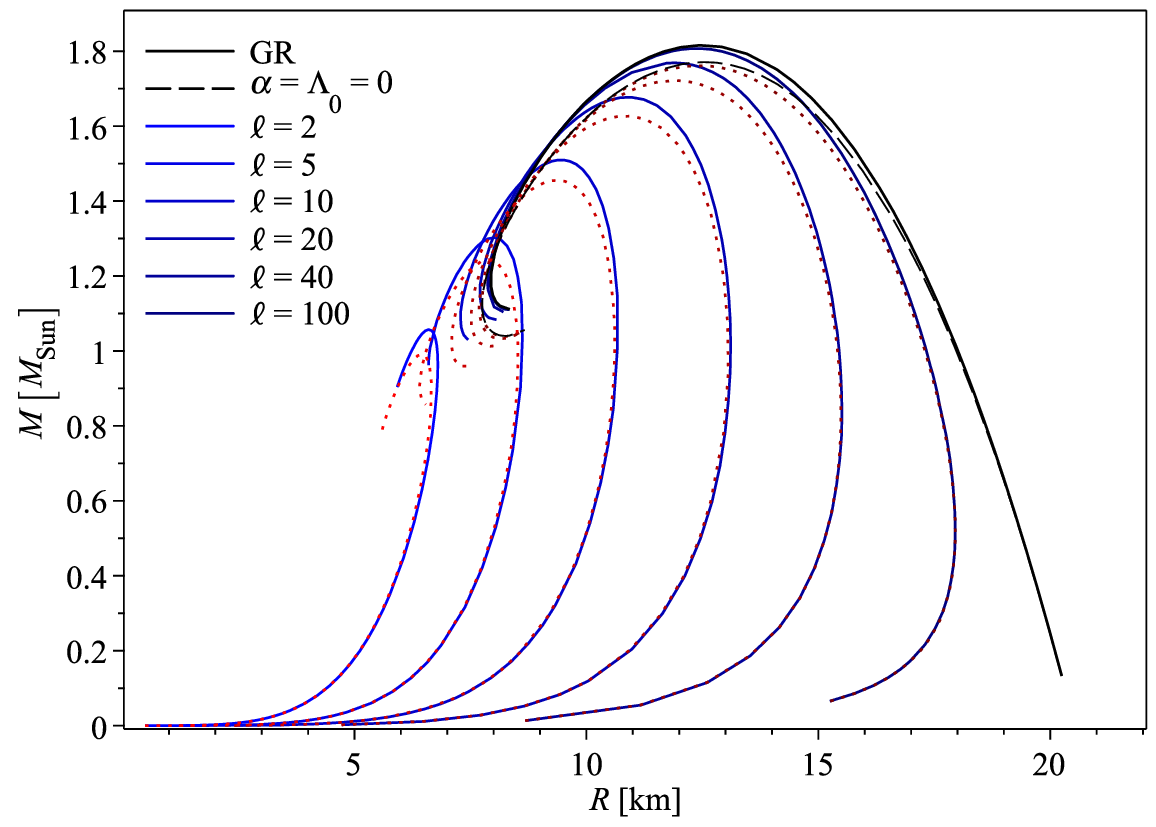}
\end{center}	
\caption{\label{MR_xi-1}
		{\em Left panel:} The mass-radius diagram in the case $\xi=-1$. Solid red curves correspond to different values of the nonminimal derivative coupling parameter $\ell=2, 5, 10, 20, 40, 100\, {\rm km}$ (from left to right). Small solid circles on the curves mark the values $\rho_{0c}=\rho_n$ and $\rho_{0c}=10\rho_n$. Thin dotted lines mark observable restrictions for mass, $1.1 M_\odot\le M_{NS}\le 2.05 M_\odot$, and radius, $10\, {\rm km}\le R_{NS}\le 14\, {\rm km}$, of the neutron star.
		{\em Right panel:} The mass-radius diagram in GR with a negative cosmological constant $\Lambda_{AdS}$. Solid blue curves correspond to different values of $\Lambda_{AdS}=-1/\ell^2$ where $\ell=2, 5, 10, 20, 40, 100\, {\rm km}$ (from left to right).
		The black solid curve corresponds to the mass-radius diagram in GR. The black dashed curve corresponds to the solution obtained earlier by Cisterna et al \cite{Rinaldi:2015} for the particular case  $\alpha=\Lambda_0=0$. All graphs are built in the range of values of the central baryonic mass density $\rho_{0c}$ from $0.1\rho_n$ to $100\rho_n$ (from bottom to up), where $\rho_n=2.5\times10^{14}\ g/cm^3$ is the nuclear density. }
\end{figure}

Results of numerical integration are given in Figs. \ref{ABRhoPsi}, \ref{MR_rho_xi-1}, \ref{MR_xi-1}. In Fig. \ref{ABRhoPsi} we demonstrate the typical behavior of the functions $A(r)$, $B(r)$, $p(r)$, and $\psi^2(r)$ inside and outside the star. A dependence of mass and radius of the star on the central baryonic mass density $\rho_{0c}$ is shown in Fig. \ref{MR_rho_xi-1}. 
More detailed information about neutron star configurations can be extracted from a mass-radius diagram. In Fig. \ref{MR_xi-1} (left panel) we present the mass-radius diagrams obtained for different values of the nonminimal derivative coupling parameter $\ell$. Also, for comparison, we present in the same plot the mass-radius diagram obtained earlier by Cisterna et al \cite{Rinaldi:2015} for the particular case  $\alpha=\Lambda_0=0$ and the mass-radius diagram for GR. One can see that in the general case the relation of mass and radius tends to that obtained for the case $\alpha=\Lambda_0=0$ in the limit of large values of $\ell$. From the other hand, the less is $\ell$ the less are mass and radius of the star comparing with the case $\alpha=\Lambda_0=0$. Note that, using the mass-radius diagram given in Fig. \ref{MR_xi-1}, one can in principle restrict possible values of the nonminimal derivative coupling parameter $\ell$. 
{
	The available observable data suggests that the most neutron stars has masses close to $1.3\!-\!1.4\, M_{\odot}$ and radii $10\!-\!14\, {\rm km}$
	\cite{JamesLattimer2012, JamesLattimer2014}. 
	While the interval of values of neutron star radii is restricted quite strict, lower and higher masses presumably can exist 
	\cite{Heinke:2006, Suleimanov:2011, Suleimanov:2016, Suleimanov:2017, Miller:2017, bookAstrNS2018, AthanasiosKottas2013, NiranjanKumar2022, Doroshenko_etal:2022}. 
	} 
As the result, applying observable restrictions for mass, $1.1 M_\odot\le M_{NS}\le 2.05 M_\odot$, and radius, $10\, {\rm km}\le R_{NS}\le 14\, {\rm km}$, we get the following estimation: $\ell\ge 10\, {\rm km}$. 

It is necessary to emphasize that mass-radius diagram in Fig. \ref{MR_xi-1} obtained for our model has an essential difference comparing with that in GR or the Cisterna et al case. Namely, the mass-radius relation corresponds to the so-called bare strange stars or quark stars (see the excellent book \cite{Book_HPY} and references therein). The main feature of bare strange stars is that their radius decreases monotonically with decreasing $M$, with $R \propto M^{1/3}$ for small enough masses of the star. Such the property of strange stars is explained with using the Bag Model for describing quark matter \cite{Book_HPY}. 
{
 Note that recently, in Ref. \cite{Doroshenko_etal:2022} the central compact object within the supernova remnant HESS J1731-347 has been interpreted as a neutron star with  the mass and radius to be $M=0.77^{+0.20}_{-0.17}M_\odot$ and $R=10.4^{+0.86}_{-0.78}$ km, respectively, based on modelling of the X-ray spectrum and a robust distance estimate from Gaia observations. This analysis implies that the central object is either the lightest neutron star known, or a ‘strange star’ with a more exotic equation of state.
}

As we will see below, the specific `strange' relation between mass and radius in our case is forming due to the negative cosmological constant $\Lambda_{AdS}$ given by Eq. (\ref{Leff}). To illustrate this fact, below we consider neutron stars in Einstein's theory of gravity without any scalar field but with a negative cosmological constant. 

\subsubsection{Bare strange stars in General Relativity with a negative cosmological constant}
Let us consider the theory of gravity with the action
\begin{equation}\label{action_AdS}
	S=\int d^4x\sqrt{-g}\,
	\left[\frac{1}{2\kappa} (R+2|\Lambda_{AdS}|)\right]+S^{(m)},
\end{equation}
where $\Lambda_{AdS}$ is a negative cosmological constant.
The field equations (\ref{fe1}), (\ref{fe2}), (\ref{eom}) reduce to the following form:
\begin{eqnarray}
	-\frac{B'}{r}+\frac{1-B}{r^2} &=& \kappa\epsilon -|\Lambda_{AdS}|,
	\\
	\frac{B A'}{r A}-\frac{1-B}{r^2} &=& \kappa p+|\Lambda_{AdS}|,
	\\
	\frac{A'}{A} &=& -\frac{2p'}{\epsilon +p}.
\end{eqnarray}
These equations can be easily rewritten in the standard form as follows
\begin{equation}
	\label{modTOV}	
	-r^2\frac{dp}{dr}=G{\cal M}\rho
	\left(1+\frac{p}{\epsilon}\right)
	\left(1+\frac{4\pi r^3}{{\cal M}}
	\left(p+\frac{c^4|\Lambda_{AdS}|}{6\pi G}\right)\right)
	\left(1-\frac{2G{\cal M}}{c^2 r} +\frac{|\Lambda_{AdS}|}{3}\,r^2\right)^{-1}.
\end{equation}
\begin{equation}
	\label{modB}
	B(r)=1-\frac{2G{\cal M}(r)}{c^2 r} +\frac{|\Lambda_{AdS}|}{3}\, r^2,
\end{equation}
\begin{equation}
	\label{modM}
	{\cal M}(r)=\int_0^r 4\pi r^2 \rho dr,
\end{equation}
where $\rho=\epsilon/c^2$, and Eq. (\ref{modTOV}) is the Tolman-Oppenheimer-Volkoff (TOV) equation modified by the $\Lambda_{AdS}$-term. Results of numerical integration of Eqs. (\ref{modTOV})--(\ref{modM}) for different values of $\Lambda_{AdS}=-1/\ell^2$ are shown in Fig. \ref{MR_xi-1} (right panel). One can see that the relation between mass and radius is closely similar to that given in the left panel of Fig. \ref{MR_xi-1} for the model with nonminimal derivative coupling in case $\xi=-1$, when  $\Lambda_{AdS}=\Lambda_0=-1/\ell^2$. Moreover, for small enough values of the central baryonic mass density $\rho_{0c}$ the mass-radius diagrams are in practice indistinguishable. Note also that in the limit of large values of $\ell$, when $\Lambda_{AdS}\to 0$, the relation between mass and radius tends to that obtained for GR, which in turn slightly differs from that obtained by  Cisterna et al \cite{Rinaldi:2015} in the case $\alpha=\Lambda_0=0$.

\subsubsection{The case $\xi\not=-1$ }
Generally, $-3<\xi<1$ and $\xi\not=-1$. 
\begin{figure}[th]
	\begin{center}
		\includegraphics[scale=0.4]{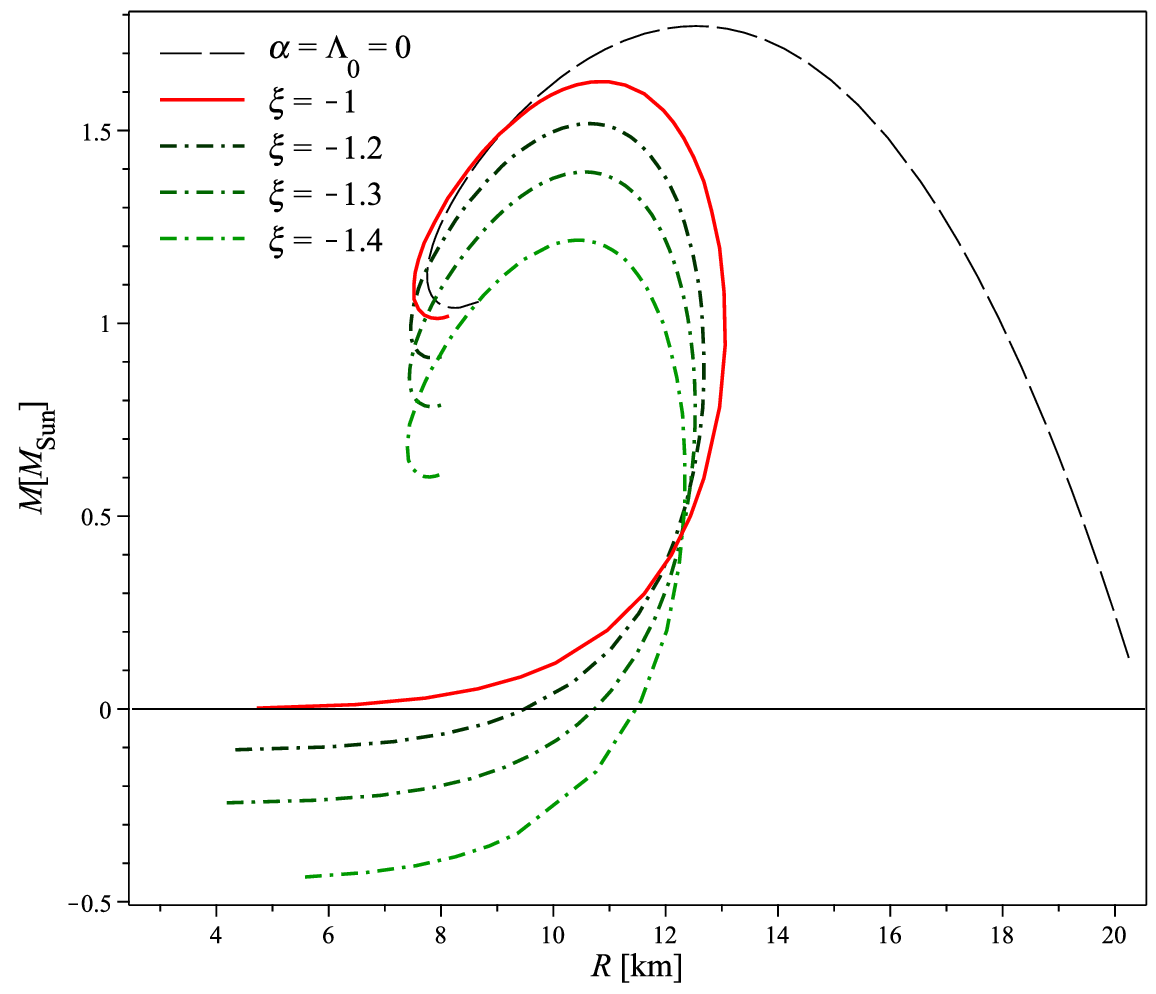}\ 
		\includegraphics[scale=0.4]{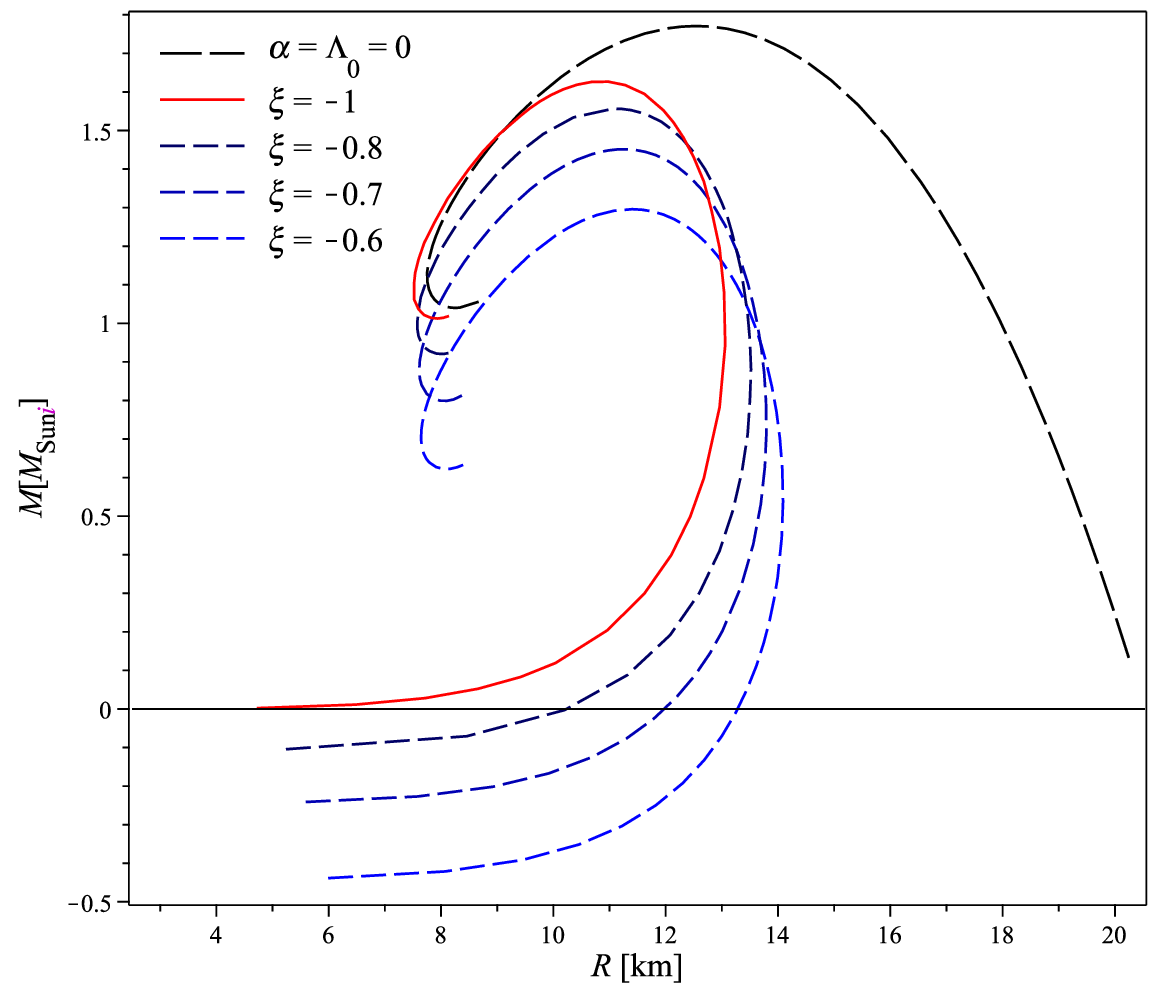}
	\end{center}	
	\caption{\label{MR_xi_not_-1}
		The mass-radius diagram in the case $\xi\not=-1$ are shown for $\ell=20\, {\rm km}$. The solid red curve corresponds to the case $\xi=-1$. The black dashed curve corresponds to the solution obtained earlier by Cisterna et al \cite{Rinaldi:2015} for the particular case  $\alpha=\Lambda_0=0$. All graphs are built in the range of values of the central baryonic mass density $\rho_{0c}$ from $0.1\rho_n$ to $100\rho_n$ (from bottom to up), where $\rho_n=2.5\times10^{14}\ g/cm^3$ is the nuclear density. 
		Left panel: Green dash-dot curves correspond to $\xi=-1.2,-1.3,-1.4$.
		Right panel: Blue dashed curves correspond to $\xi=-0.8,-0.7,-0.6$.
		}
\end{figure}
Mass-radius diagrams for various values of $\xi\not=-1$ are shown in Fig. \ref{MR_xi_not_-1}. One can see that in comparison with the case $\xi=-1$  mass-radius diagrams are shifted down and left in case $\xi<-1$, and down and right in case $\xi>-1$. It is also necessary to note that mass-radius diagrams are shifted in the region of negative masses. Of course the baryonic mass of the star defined as
\be
M_0=\int_0^R 4\pi B^{-1/2} \rho_0\, dr,
\ee 
remains to be positive. However, it turns out that the asymptotic anti-de Sitter mass given by Eqs. (\ref{A_AdS})-(\ref{Leff}) becomes negative for some values the central baryonic mass density $\rho_{0c}$.

{
\section{Stability of anti-de Sitter neutron stars} \label{sec5}
An investigation of stability of solutions describing relativistic compact objects is an important and rather complicated problem which needs a separate study.\footnote{We thank the anonymous referee who brought this problem to our attention and led us to investigate stability of anti-de Sitter neutron stars.} In the most of problems we are primarily interested in stable solutions. Might solutions have been unstable, their astrophysical importance would probably have been depreciated. However this does not make the study of such solutions useless, unstable solutions with a very long unstability timescale also might still be relevant in some context.

A method to study stability of general relativistic black holes has been derived in the pioneering works by Regge and Wheeler \cite{ReggeWheeler} and Zerilli \cite{Zerilli}. Later, this approach has been widely used to investigate the stability of black holes and neutron stars in various modified theories of gravity. 
In particular, perturbations and quasi-normal modes of static and spherically symmetric black holes and neutron stars in Horndeski theory have been intensively studied in Refs.
\cite{
	KobMotSuy:I2012, 
	KobMotSuy:II2014, 
	OgaKobSuy:2016, 
	TakSui:2017, 
	TatFer:2018,
	KasKimSatTsu:2020, 
	TomKob:2021, 
	LanHouRou:2021, 
	MinTakTsu:2022a, 
	MinTakTsu:2022b, 
	KasTsu:2021,
	KasTsu:2022, 
	KhoNouTroWong:2022, 
	MinTsu:2022}.
Shortly, the essence of Regge-Wheeler method consists in the following (one can find details in Refs. \cite{ReggeWheeler, KobMotSuy:I2012, KobMotSuy:II2014, KasTsu:2022}): To study the stability of relativistic compact objects (black holes or neutron stars), one considers metric perturbations $h_{\mu\nu}$, such that
$$g_{\mu\nu} = \bar g_{\mu\nu} + h_{\mu\nu},$$ 
where $\bar g_{\mu\nu}$ is the background metric of the spherically symmetric and static spacetime. Analogously, scalar and matter fields are perturbed. All perturbations are divided into scalar, vector, and tensor ones, expanded into spherical harmonics, and then odd- and even-parity modes are analyzed separately. In such the approach, stability conditions could be considered as a positivity of squared propagation speeds of odd- and even-parity perturbations along the radial and angular directions.  

Stability of ``stealth'' neutron stars in the theory of gravity with a nonminimal scalar-derivative coupling with a vanishing ``bare'' cosmological constant, $\Lambda_0=0$, and a vanishing standard kinetic term, $\alpha=0$, constructed by Cisterna et al. \cite{Rinaldi:2015},
%
has been studied in details in \cite{KasTsu:2021, KasTsu:2022}. The main result obtained in these works can be shortly expressed as follows: The Laplacian instability associated with negative values of $c^2_{\Omega-}$ is
always present in case ${\cal C} < 1/3$, where $c^2_{\Omega-}$ is the squared propagation speeds of even-parity perturbations of the scalar field in the angular direction around the surface of star, and $\cal C$ is the compactness parameter, ${\cal C}=GM_{NS}/R_{NS}$. Even for ${\cal C} > 1/3$, the neutron star solutions are subject to ghost instabilities. The final conclusion made in Ref. \cite{KasTsu:2021} is that there are no stable neutron star configurations in the derivative coupling theory without a standard kinetic term, including both relativistic and nonrelativistic compact objects.

In this work we have constructed neutron star configurations in the derivative coupling theory in the general case with $\Lambda_0\not=0$ (nonzero bare cosmological term) and $\alpha\not=0$ (nonzero standard kinetic term). The full results of the stability analysis of the found configurations are very extensive and will be represented in a separate publication. However, here we are shortly touching some important points. Following Refs. \cite{KasTsu:2021, KasTsu:2022}, we have found values of the squared propagation speeds of odd- and even-parity perturbations. The values crucial for the stability analysis are the following: $c_r$ and $c_\Omega$ are speeds of the only propagating degree of freedom in odd-parity sector in radial and angular directions, respectively; $c_{r3}$ is the propagation speed of scalar-field even-parity perturbations $\delta\phi$ propagating the radial direction; $c_{\Omega-}$ is the propagation speed of the scalar-field even-parity perturbations $\delta\phi$ propagating in the angular direction;  $c_{\Omega+}$ is the propagation speed of even-parity perturbations $\psi$ propagating in the radial direction, which corresponds to the degree of freedom arising from the metric perturbations. A necessary and sufficient condition for the absence of Laplacian instability is the non-negativity of the squared speeds of perturbations:
\be
c^2_r \ge 0, \quad
c^2_\Omega \ge 0, \quad
c^2_{r3} \ge 0, \quad
c^2_{\Omega\pm} \ge 0.
\ee
Besides these conditions, there is the additional one, ${\cal K}>0$, which can be regarded as the no-ghost condition in the presence of matter; here $\cal K$ is a definite expression constructed from unperturbed values of metric components and scalar field, and depending on $\rho+P$  (see \cite{KasTsu:2021, KasTsu:2022}). 

We have studied in details the stability of neutron star configurations in the theory of gravity with nonminimal derivative coupling and found that there exists a wide class of model parameters for which all stability conditions are fulfilled. In particular, in Figs. \ref{domain1} and \ref{domain2} we demonstrate examples of star configurations for which all squared speeds of perturbations, $c^2_r$, $c^2_\Omega$, $c^2_{r3}$, $c^2_{\Omega\pm}$, and also $\cal K$ are positive. The examples are given for $\rho_{0c}=10^{15}\ {\rm g/cm^3}$ (central baryonic mass density) and $\ell=10\ {\rm km}$ (characteristic scale of the nonminimal derivative coupling). In this case the stability conditions are realized within two domains of the parameter $\xi$: (i) $\xi\in(-1.67,-1.18)$; (ii) $\xi\in(-0.3,0.08)$. Note additionally that $r^2_\Omega=1$ and ${\cal K}>0$ within these domains, hence, in principle, the neutron star configurations with such the set of parameters are free from the Laplacian and ghost instabilities. Note also that the value $\xi=-1$ lays outside these domains. For comparison, we also include into plots \ref{domain1} graphs of squared speeds corresponding to the case $\xi=-1$. It is clearly seen that $c^2_{\Omega-}$  takes negative values inside the star, that is the neutron star configurations with $\xi=-1$ are unstable. The full stability analysis of anti-de Sitter neutron stars in the theory of gravity with nonminimal derivative coupling will be given in \cite{inprep}.

\begin{figure}[th]
	\begin{center}
		\includegraphics[scale=0.38]{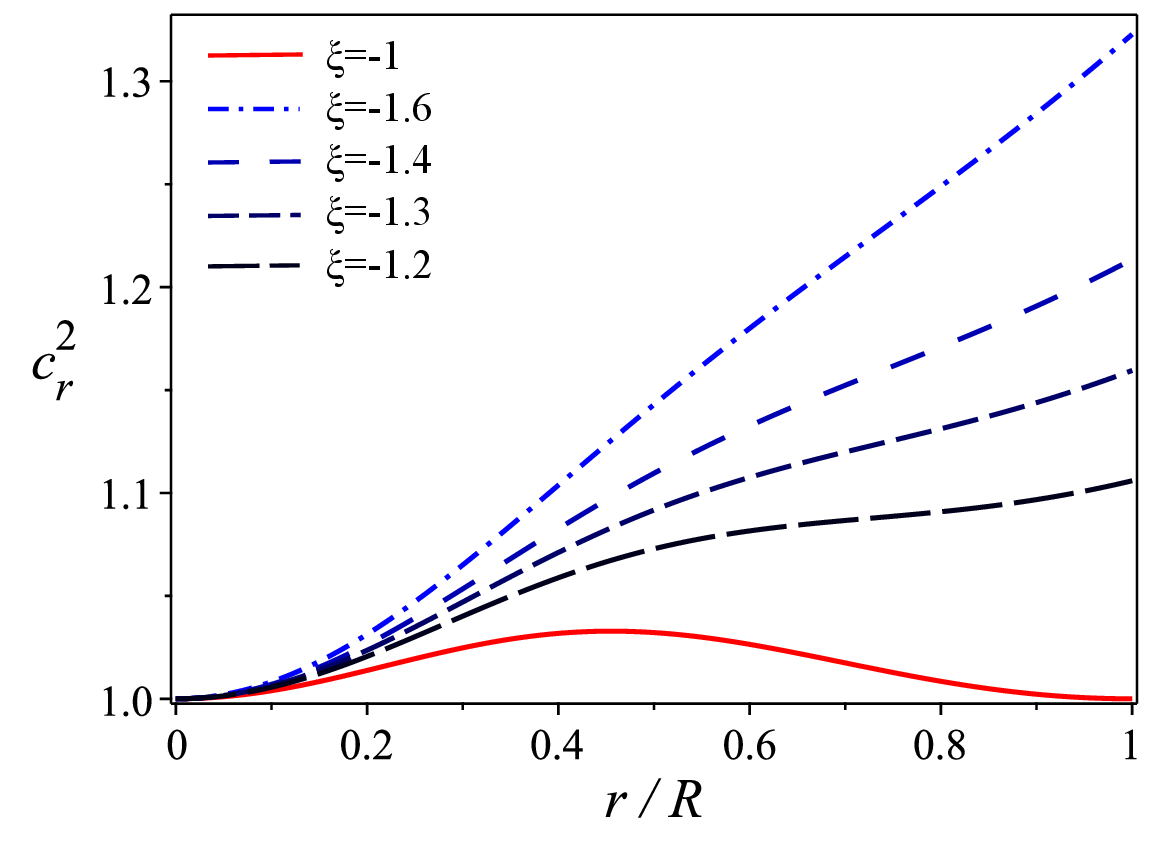}\
		\includegraphics[scale=0.38]{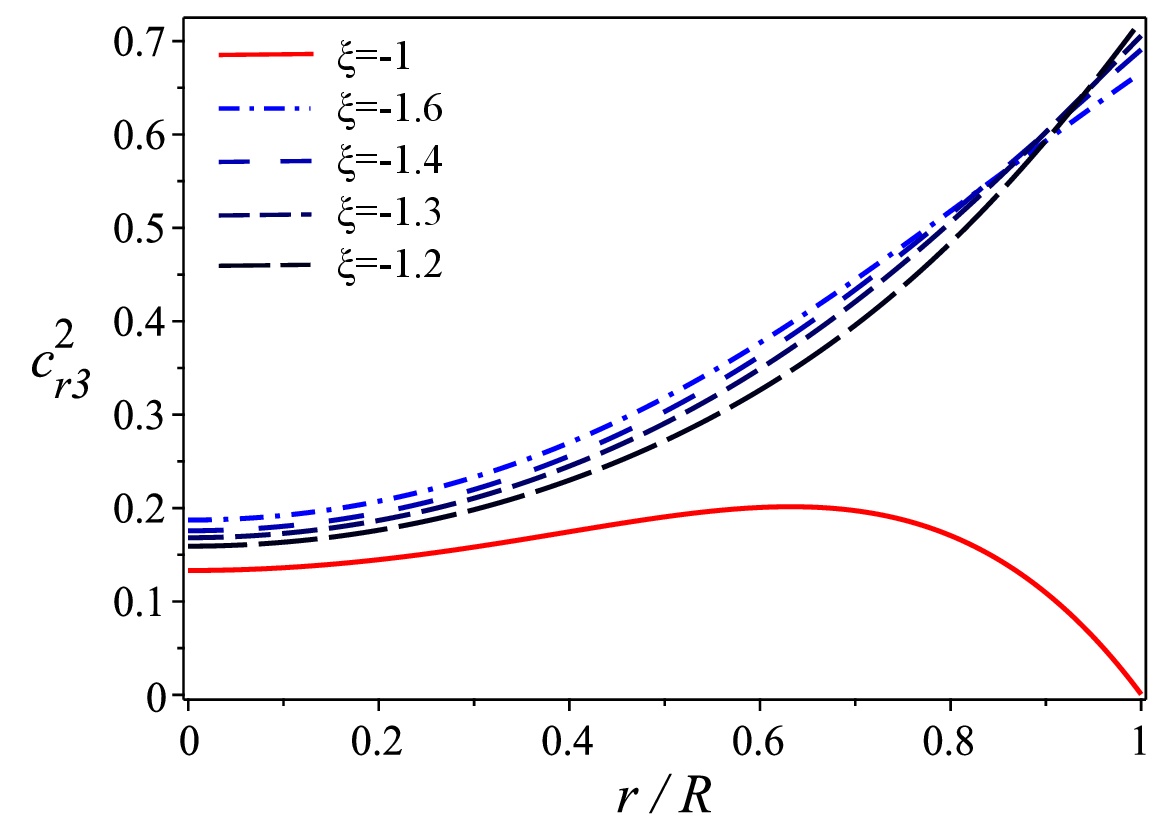}\\ 
		\includegraphics[scale=0.38]{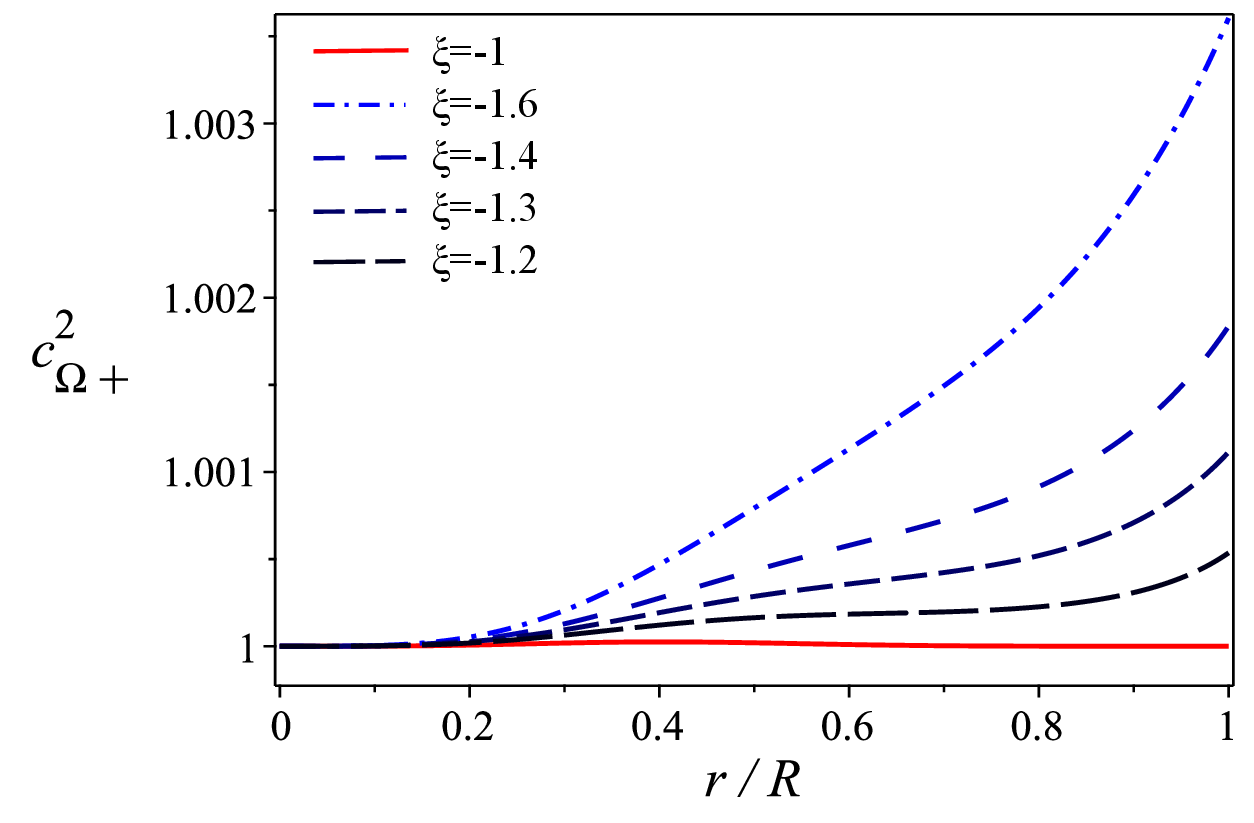}\
		\includegraphics[scale=0.38]{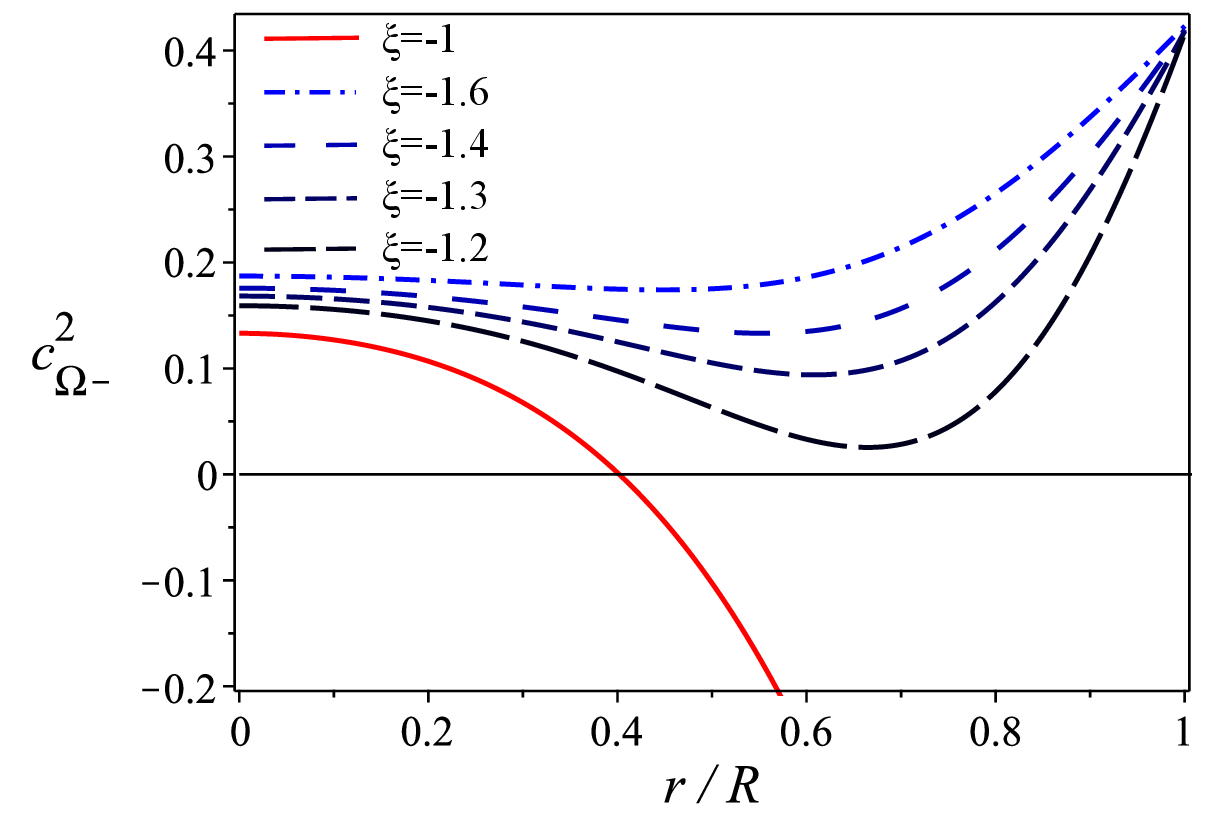}
	\end{center}	
	\caption{\label{domain1}
		Graphs of squared speeds of  perturbations $c^2_r$, $c^2_{r3}$, $c^2_{\Omega\pm}$ depending on a reduced radial coordinate $r/R$, where $R$ is the radius of a star. Blue curves correspond to the domain $\xi\in(-1.67,-1.18)$: $\xi=-1.6; -1.4; -1.3; -1.2$. The red solid curve corresponds to $\xi=-1$.
	}
\end{figure}
%
\begin{figure}[th]
	\begin{center}
		\includegraphics[scale=0.38]{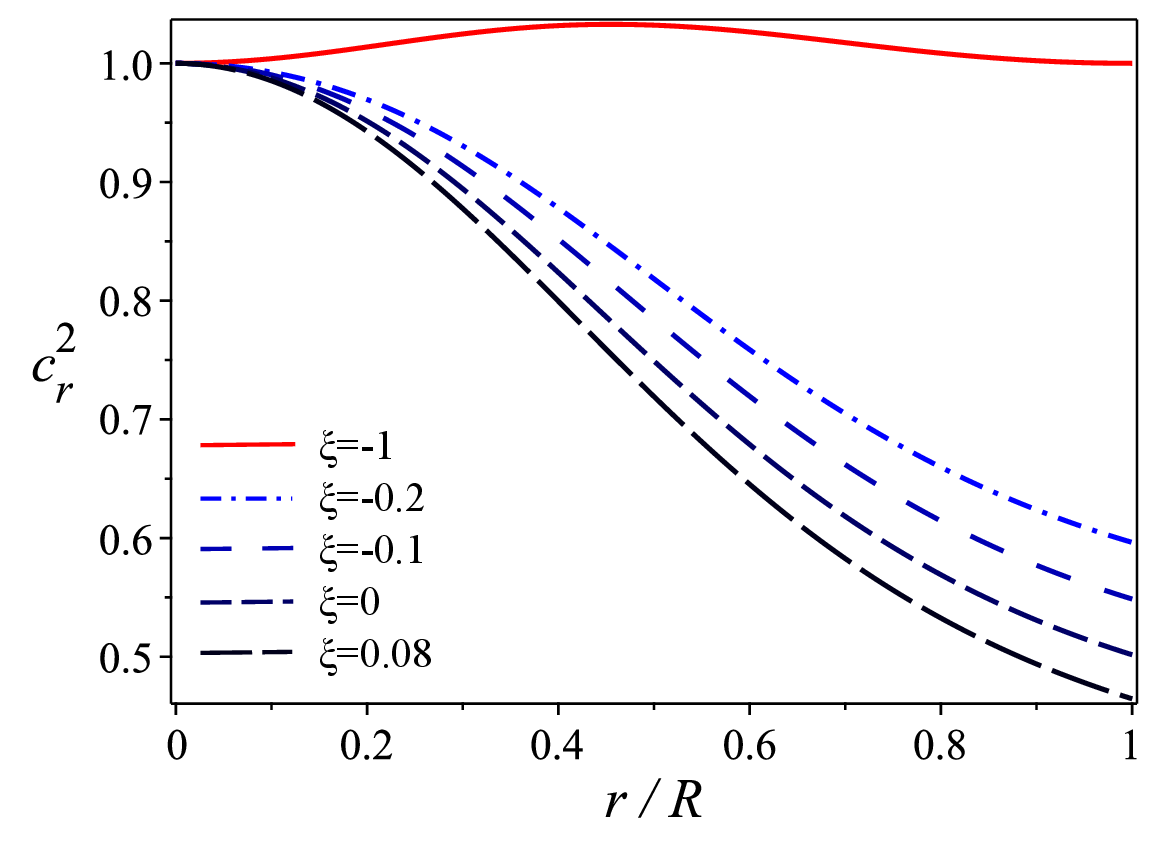}\
		\includegraphics[scale=0.38]{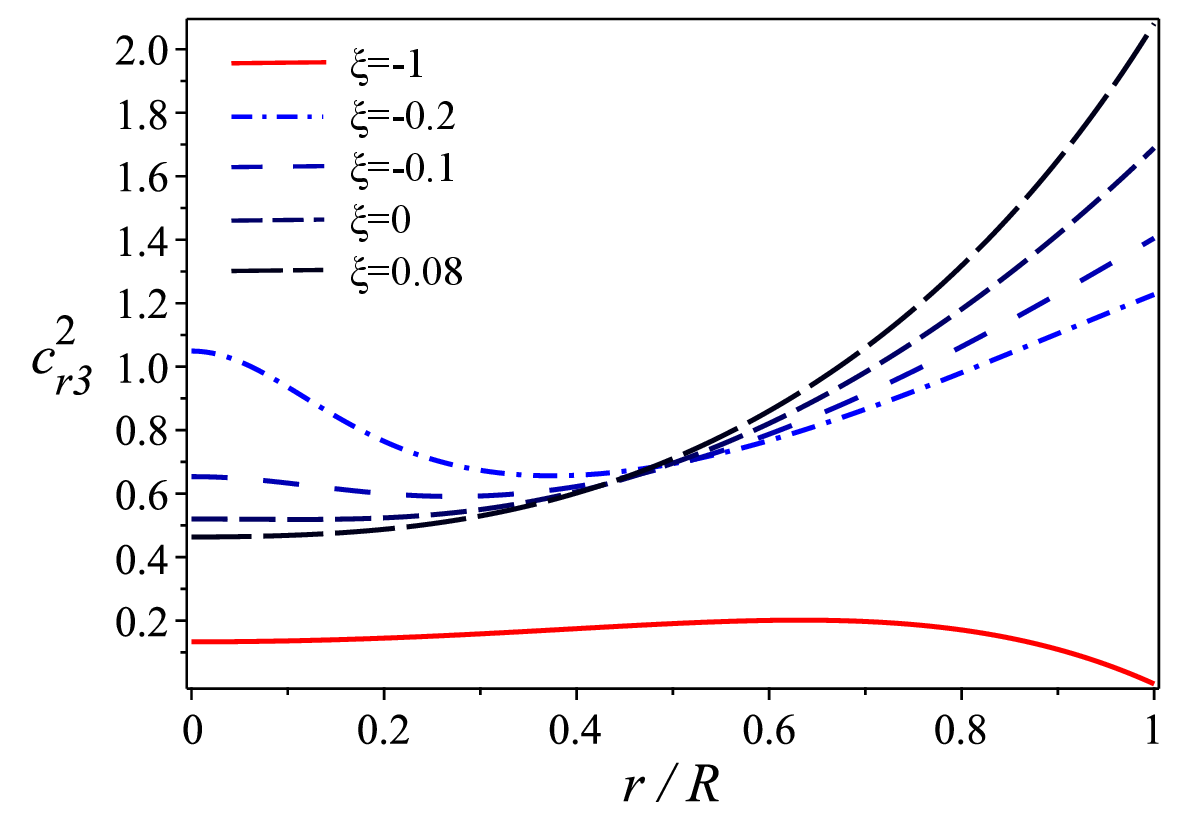}\\ 
		\includegraphics[scale=0.38]{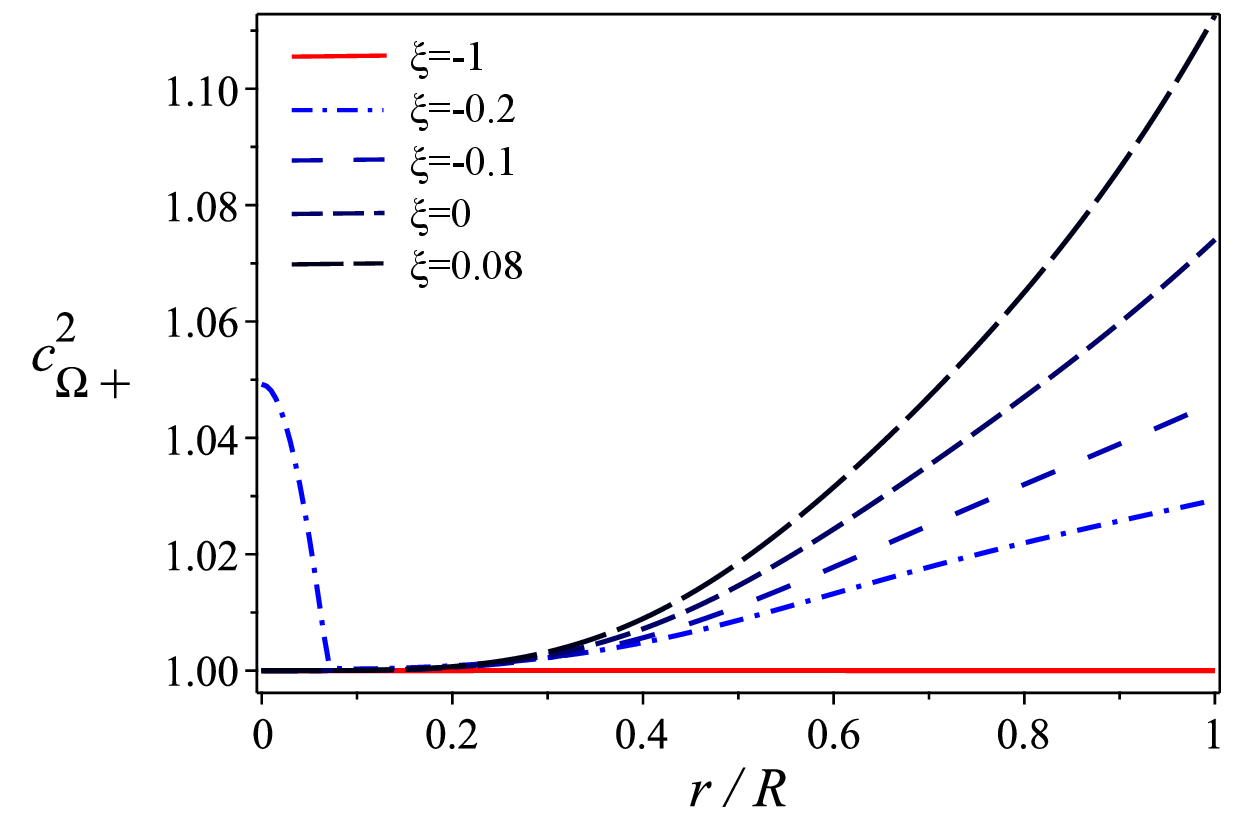}\
		\includegraphics[scale=0.38]{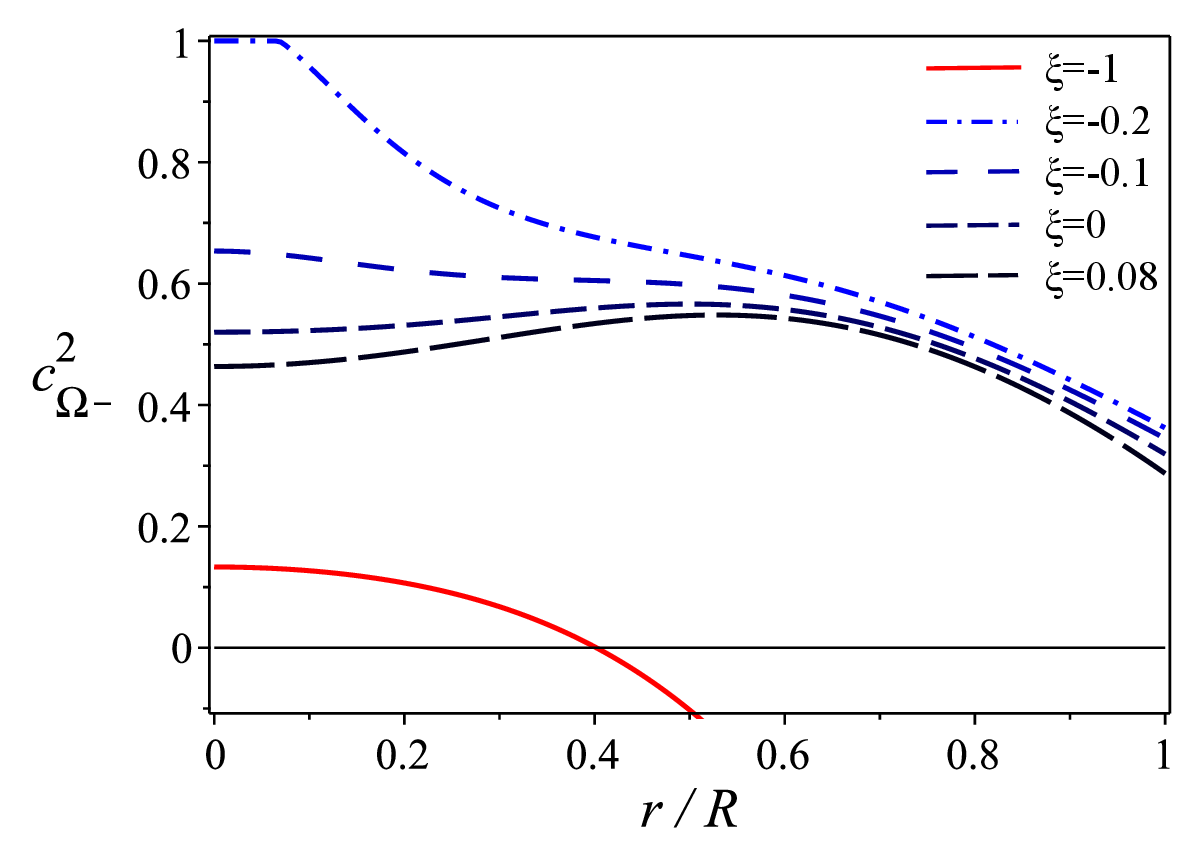}
	\end{center}	
	\caption{\label{domain2}
		Graphs of squared speeds of perturbations $c^2_r$, $c^2_{r3}$, $c^2_{\Omega\pm}$ depending on a reduced radial coordinate $r/R$, where $R$ is the radius of a star. Blue curves correspond to the domain $\xi\in(-0.3,0.08)$: $\xi=-0.2; -0.1; 0; 0.08$. The red solid curve corresponds to $\xi=-1$.
	}
\end{figure}
%

} 
 
\section{Summary} \label{Summary}
In this paper, we investigated neutron star configurations in the scalar-tensor theory of gravity with the non-minimal derivative coupling of a scalar field to curvature (\ref{action}), which is a subclass of Horndeski gravity. Neutron stars in this model were studied in a series of papers \cite{Rinaldi:2015, Rinaldi:2016, Silva:2016, Maselli:2016, Eickhoff:2018} 
for the special case with a vanishing ``bare'' cosmological constant, $\Lambda_0=0$, and a vanishing standard kinetic term, $\alpha=0$. 
This special case is of interest because it admits so-called stealth configuration, i.e. vacuum configuration with nontrivial scalar field and the Schwarzschild metric. However, generally one has $\Lambda_0\not=0$ and $\alpha\not=0$, and in this case a vacuum configuration is represented as an asymptotically anti-de Sitter (AdS) black hole solution with the nontrivial scalar field \cite{Rinaldi:2012, Minamitsuji:2013, Anabalon:2014, Babichev:2014, Kobayashi:2014, Babichev:2015}. 
Our analysis revealed that the effective cosmological constant is expressed in terms of model parameters as follows 
$$ 
\Lambda_{AdS}=-\,\frac{1-\xi}{3+\xi}\,\frac{1}{\ell^2},
$$
where $\ell$ is a characteristic length of nonminimal derivative coupling, and $\xi=\Lambda_0 \ell^2$ is a dimensionless parameter connecting with the ``bare'' cosmological constant $\Lambda_0$. We found that $-3<\xi<1$, and so $\Lambda_{AdS}$ is negative for all values of $\Lambda_0$ including zero.

{
	Analyzing the boundary conditions, we obtained that the scalar charge of a neutron star configuration is inevitably equal to zero, $Q=0$. This means that possible stellar configuration has no scalar hairs in a total consistence with the result proved in \cite{Lehebel:2017fag}.
}

An internal structure of neutron stars was explored numerically for different sets of model parameters $\xi$ and $\ell$ and for matter considered as a perfect fluid with the polytropic equation of state (\ref{eos}) with $\Gamma=2$ and $K=1.79\times 10^5\, {\rm cgs}$. It turned out that resulting diagrams describing the relation between mass and radius of the star essentially differ from those obtained in GR or the particular model with $\alpha=\Lambda_0=0$ considered by Cisterna et al \cite{Rinaldi:2015}. Instead, the mass-radius diagrams are similar to those obtained for so-called bare strange stars when a star radius decreases monotonically with decreasing mass. 
Our analysis shown that such the specific relation between mass and radius is forming due to appearance of the effective negative cosmological constant $\Lambda_{AdS}$. To illustrate this fact, we considered neutron stars in Einstein's theory of gravity without any scalar field but with a negative cosmological constant. 

{
Using the Regge-Thorne method we studied in details the stability of obtained neutron star configurations and found that there exists a wide class of model parameters for which all stability conditions are fulfilled. In particular, in Figs. \ref{domain1} and \ref{domain2} we demonstrated examples of star configurations for which all squared speeds of perturbations, $c^2_r$, $c^2_\Omega$, $c^2_{r3}$, $c^2_{\Omega\pm}$, and also $\cal K$ are positive, and so such the  configurations are free from the Laplacian and ghost instabilities. The full stability analysis of anti-de Sitter neutron stars in the theory of gravity with nonminimal derivative coupling is rather extensive and will be given in a separate publication \cite{inprep}.
}

It is worth also noticing that a theoretical analysis of neutron star configurations in the theory of gravity with the nonminimal derivative coupling and comparing the theoretical results with observable data could, in principle, provide restrictions on possible values of the parameter of nonminimal coupling $\ell$. For example, in the case $\xi=-1$ we found that $\ell\ge10\, {\rm km}$. Of course, in order to find more strict constraints for $\ell$ one has to consider more realistic models of neutron stars, using more realistic equations of state, taking into account a rotation of the star, etc. Such investigations are planned in the future.  

\section*{Acknowledgments}
The authors acknowledge very helpful and fruitful discussions with Valery Suleimanov. This work is supported by the RSF grant No. 21-12-00130 and partially carried out in accordance with the Strategic Academic Leadership Program "Priority 2030" of the Kazan Federal University.



\begin{thebibliography}{99}
\bibitem{Will}
C. M. Will, The Confrontation between General Relativity and Experiment, Living Rev. Rel. {\bf 17}, 4 (2014); arXiv:1403.7377 [gr-qc].

\bibitem{Review_Salvatore:2011} 
S. Capozziello, M. De Laurentis, Extended Theories of Gravity, Phys. Rep. {\bf 509}, Issues 4–5, pp. 167-321 (2011).

\bibitem{Review_Clifton_etal}
T. Clifton, P. G. Ferreira, A. Padilla, and C. Skordis, Modified Gravity and Cosmology, Phys. Rept. {\bf 513}, 1–189 (2012).

\bibitem{Review_ModGrav:2013} 
R. Myrzakulov, L. Sebastiani, S. Zerbini, Some aspects of generalized modified gravity models, 	Int. J. Mod. Phys. D {\bf 22}, no.8, 1330017 (2013).

\bibitem{Review_Berti_etal}
E. Berti et al., Testing General Relativity with Present and Future Astrophysical Observations, Class. Quant. Grav. {\bf 32}, 243001 (2015).

\bibitem{Review_Nojiri:2017} 
S. Nojiri, S. D. Odintsov, V. K. Oikonomou, Modified gravity theories on a nutshell: Inflation, bounce and late-time evolution, 	Phys. Rep.  {\bf 692}, pp. 1-104 (2017).

\bibitem{Review_Langlois:2019} 
D. Langlois, Dark energy and modified gravity in degenerate higher-order scalar–tensor (DHOST) theories: A review, Int. J. Mod. Phys. D {\bf 28}, no. 05, 1942006 (2019).

\bibitem{Horndeski}
G. W. Horndeski, Second-order scalar-tensor field equations in a four-dimensional space, Int. J. Theor. Phys. {\bf 10}, 363–384 (1974).

\bibitem{Kobayashi:2011} 
T. Kobayashi, M. Yamaguchi and J. Yokoyama, 
Prog. Theor. Phys. {\bf 126}, 511 (2011).

\bibitem{Book_HPY}
P. Haensel, A. Y. Potekhin, and D. G. Yakovlev, Neutron Stars 1, Springrer (2007)

\bibitem{LattimerPrakash} 
J.M. Lattimer, M. Prakash, The Physics of Neutron Stars, Science, {\bf 304},  5670, 536-542 (2004).


\bibitem{PageReddy:2006} 
D. Page, S. Reddy, Dense Matter in Compact Stars: Theoretical Developments and Observational Constraints, Ann. Rev. Nucl. Part. Sci. {\bf 56}, no.1, pp.327-374 (2006). 

\bibitem{Schmitt} 
A. Schmitt, Dense matter in compact stars: A pedagogical introduction, Lecture Notes in Physics, {\bf 811}, 1-111 (2010).




\bibitem{Babichev:2009} 
E. Babichev, D. Langlois, 
Relativistic stars in $f(R)$ gravity,  
Phys. Rev. D {\bf 80}, 121501 (2009). 

\bibitem{Babichev:2010} 
E. Babichev, D. Langlois, 
Relativistic stars in $f(R)$ and scalar-tensor theories, 
Phys. Rev. D {\bf 81}, 124051 (2010).

\bibitem{Cooney:2010} 
A. Cooney, S. DeDeo, D. Psaltis, 
Neutron stars in $f(R)$ gravity with perturbative constraints, 
Phys. Rev. D {\bf 82}, 064033 (2010).

\bibitem{Orellana:2013} 
M. Orellana, F. García, F.A. Teppa Pannia, et al. 
Structure of neutron stars in R-squared gravity. 
Gen. Relativ Gravit {\bf 45}, 771–783 (2013). 

\bibitem{Ryotaro:2019} 
R. Kase, S. Tsujikawa, Neutron stars in $f(R)$ gravity and scalar-tensor theories, J. Cosmol. Astropart. Phys. {\bf 2019}, no. 09, 054 (2019).

{
\bibitem{Astashenok:2021peo}
A.~V.~Astashenok, S.~Capozziello, S.~D.~Odintsov and V.~K.~Oikonomou,
Causal limit of neutron star maximum mass in $f(R)$ gravity in view
of GW190814, Phys. Lett. B \textbf{816}, 136222 (2021).

\bibitem{Astashenok:2020qds}
A.~V.~Astashenok, S.~Capozziello, S.~D.~Odintsov and V.~K.~Oikonomou,
Extended Gravity Description for the GW190814 Supermassive Neutron
Star, Phys. Lett. B \textbf{811}, 135910 (2020).

\bibitem{Astashenok:2021xpm}
A.~V.~Astashenok, S.~Capozziello, S.~D.~Odintsov and V.~K.~Oikonomou,
Novel stellar astrophysics from extended gravity,
EPL \textbf{134}, no.5, 59001 (2021).

\bibitem{Astashenok:2021btj}
A.~V.~Astashenok, S.~Capozziello, S.~D.~Odintsov and V.~K.~Oikonomou,
Maximum baryon masses for static neutron stars in f(R) gravity,
EPL \textbf{136}, no.5, 59001 (2021).
}


%
%

\bibitem{Moraes:2016} 
P.H.R.S. Moraes, J. D. Arbañil, M. Malheiro, Stellar equilibrium configurations of compact stars in $f(R,T)$ gravity, J. Cosmol. Astropart. Phys. {\bf 06}, 005 (2016).

\bibitem{Pace:2017} 
M. Pace, J. L. Said, A Perturbative Approach to Neutron Stars in $f(R,T)$-Gravity, Eur. Phys. J. C {\bf 77}, 283 (2017). 

\bibitem{Mathew:2020} 
A. Mathew, M. Shafeeque, M. K. Nandy, Stellar structure of quark stars in a modified Starobinsky gravity, Eur. Phys. J. C {\bf 80}, 615 (2020). 

\bibitem{Carvalho2020} 
G.A. Carvalho, P.H.R.S. Moraes,  S.I. dos Santos, et al. Hydrostatic equilibrium configurations of neutron stars in a non-minimal geometry-matter coupling theory of gravity, 
Eur. Phys. J. C {\bf 80}, 483 (2020). 

\bibitem{Pretel:2021} 
J. M. Z. Pretel, S. E. Jorás, R. R. R. Reis,  J. D. V. Arbanil, Neutron stars in $f(R,T)$ gravity with conserved energy-momentum tensor: Hydrostatic equilibrium and asteroseismology,
JCAP {\bf 08}, 055 (2021).

\bibitem{Ilijic:2018} 
S. Ilijic, M. Sossich, Compact stars in $f(T)$ extended theory of gravity, Phys. Rev. D {\bf 98}, 064047 (2018).

\bibitem{Lin:2022} 
Rui-Hui Lin, Xiao-Ning Chen, Xiang-Hua Zhai, Realistic neutron star models in $f(T)$ gravity, Eur. Phys. J. C {\bf 82}, 308 (2022). 

\bibitem{Pani:2011}
P. Pani, E. Berti, V. Cardoso, J. Read,
Compact stars in alternative theories of gravity. Einstein-Dilaton-Gauss-Bonnet gravity,
Phys. Rev. D {\bf 84}, 104035 (2011). 
\bibitem{Doneva:2019} 
D. D. Doneva, S. S. Yazadjiev, Neutron star solutions with curvature induced scalarization in the extended Gauss-Bonnet scalar-tensor theories, J. Cosmol. Astropart. Phys. {\bf 1804}, 011 (2018). 

\bibitem{Horbatsch:2011} 
M.W. Horbatsch, C.P. Burgess, Semi-Analytic Stellar Structure in Scalar-Tensor Gravity, J. Cosmol. Astropart. Phys. {\bf 1108}, 027 (2011). 

\bibitem{Doneva:2020} 
D. D. Doneva, S. S. Yazadjiev, Non-topological spontaneously scalarized neutron stars in tensor-multi-scalar theories of gravity, Phys. Rev. D {\bf 101}, 104010 (2020). 

\bibitem{Raissa} 
R. F. P. Mendes, N. Ortiz, N. Stergioulas, 
Nonlinear dynamics of oscillating neutron stars in scalar-tensor gravity, Phys. Rev. D {\bf 104}, 104036 (2021).

\bibitem{OdintsovOikonomou} 
S. D. Odintsov, V. K. Oikonomou, Neutron Stars in Scalar-tensor Gravity with Higgs Scalar Potential, 	arXiv:2104.01982 (2022).

\bibitem{Rosca} 
Rosca-Mead R, Moore CJ, Sperhake U, Agathos M, Gerosa D. 
Structure of Neutron Stars in Massive Scalar-Tensor Gravity. 
Symmetry. 2020; 12(9):1384. 

\bibitem{Oliveira:2015} 
A. M. Oliveira, H. E. S. Velten, J. C. Fabris, L. Casarini, Neutron Stars in Rastall Gravity, Phys. Rev. D {\bf 92}, 044020 (2015). 

\bibitem{HarkoLoboSushkov} 
T. Harko, F. S. N. Lobo, M. K. Mak, and S. V. Sushkov,
Structure of neutron, quark, and exotic stars in Eddington-inspired Born-Infeld gravity, 
Phys. Rev. D {\bf 88}, 044032 (2013).

\bibitem{Greenwald:2010} 
J. Greenwald, A. Papazoglou, A. Wang, Black holes and stars in Horava-Lifshitz theory with projectability condition, Phys.Rev. D {\bf 81}, 084046 (2010).

\bibitem{Olmo:2020} 
G. J. Olmo, D. Rubiera-Garcia, A. Wojnar, Stellar structure models in modified theories of gravity: Lessons and challenges, Phys. Rep. {\bf 876}, pp. 1-75 (2020). 

\bibitem{Rinaldi:2015} 
A. Cisterna, T. Delsate, M. Rinaldi, 
Neutron stars in general second order scalar-tensor theory: The case of nonminimal derivative coupling, 
Phys. Rev. D {\bf 92}, 044050 (2015).

\bibitem{Rinaldi:2016} 
A. Cisterna, T. Delsate, L. Ducobu, and M. Rinaldi, 
Slowly rotating neutron stars in the nonminimal derivative coupling sector of Horndeski gravity, 
Phys. Rev. D {\bf 93}, 084046 (2016).


\bibitem{Maselli:2016} 
A. Maselli, H. O. Silva, M. Minamitsuji, E. Berti, 
Neutron stars in Horndeski gravity, 
Phys. Rev. D {\bf 93}, 124056 (2016).


\bibitem{Silva:2016} 
H. O. Silva, A. Maselli, M. Minamitsuji, E. Berti, 
Compact objects in Horndeski gravity, 
Int. J. Mod. Phys. D {\bf 25}, 09, 1641006 (2016).  

%

\bibitem{Eickhoff:2018} 
J. L. Blázquez-Salcedo, K. Eickhoff, 
Axial quasinormal modes of static neutron stars in the nonminimal derivative coupling sector of Horndeski gravity: Spectrum and universal relations for realistic equations of state, 
Phys. Rev. D {\bf 97}, 104002 (2018).


{
	\bibitem{Lehebel:2017fag}
	A. Leh\'ebel, E. Babichev, C. Charmousis, 
	A no-hair theorem for stars in Horndeski theories,
	JCAP {\bf 07}, 037 (2017).
}

%
%







\bibitem{Sushkov:2009}
S. V. Sushkov, 
Exact cosmological solutions with nonminimal derivative coupling, 
Phys. Rev. D {\bf 80}, 103505  (2009).

\bibitem{Sushkov:2010}
E. N. Saridakis, S. V. Sushkov, 
Quintessence and phantom cosmology with non-minimal derivative coupling, 
Phys. Rev. D {\bf 81}, 083510 (2010).

\bibitem{Sushkov:2012}
S. Sushkov, 
Realistic cosmological scenario with non-minimal kinetic coupling, 
Phys. Rev. D {\bf 85}, 123520 (2012). 

\bibitem{Sushkov:2016}
A. A. Starobinsky, S. V. Sushkov and M. S. Volkov,
The screening Horndeski cosmologies, 
JCAP {06}, 007 (2016).

\bibitem{Sushkov:2020}
A. A. Starobinsky, S. V. Sushkov and M. S. Volkov,
Anisotropy screening in Horndeski cosmologies,
Phys. Rev. D {\bf 101}, 6, 064039 (2020)

\bibitem{Rinaldi:2012} 
M. Rinaldi, 
Black holes with non-minimal derivative coupling, 
Phys. Rev. D {\bf 86}, 084048  (2012).

\bibitem{Minamitsuji:2013} 
M. Minamitsuji, 
Solutions in the scalar-tensor theory with nonminimal derivative coupling, Phys. Rev. D {\bf 89}, 064017 (2014).

\bibitem{Anabalon:2014} 
A. Anabalon, A. Cisterna and J. Oliva,
Asymptotically locally AdS and flat black holes in Horndeski theory,
Phys. Rev. D {\bf 89}, 084050 (2014).

\bibitem{Babichev:2014}
E. Babichev, C. Charmousis,
Dressing a black hole with a time-dependent Galileon,
JHEP {\bf 08}, 106  (2014).

\bibitem{Kobayashi:2014}
T. Kobayashi and N. Tanahashi, 
Exact black hole solutions in shift symmetric scalar–tensor theories,
PTEP {\bf 2014}, 073E02 (2014). 

\bibitem{Babichev:2015}
E. Babichev, C. Charmousis, M. Hassaine,
Charged Galileon black holes,
JCAP {\bf 05}, 031 (2015).

\bibitem{Sushkov:2012b}
S. V. Sushkov, R. Korolev,
Scalar wormholes with nonminimal derivative coupling,
Class. Quant. Grav. {\bf 29}, 085008 (2012).

\bibitem{Sushkov:2014}
R. V. Korolev,, S. V. Sushkov,
Exact wormhole solutions with nonminimal kinetic coupling,
Phys. Rev. D {\bf 90}, 124025 (2014).


\bibitem{Tooper:1964} R.F. Tooper, General Relativistic Polytropic Fluid Spheres. Astrophysical Journal, {\bf 140}, 434 (1964).

\bibitem{Tooper:1965} 
R.F. Tooper, Adiabatic Fluid Spheres in General Relativity. The Astrophysical Journal, {\bf 142}, 1541-1562 (1965).

\bibitem{Tooper:1966} 
R.F. Tooper, The "standard Model" for Massive Stars in General Relativity.  Astrophysical Journal, {\bf 143}, 465 (1966).

\bibitem{Chandrasekhar}
S. Chandrasekhar, Introduction to the study of stellar structure (The University of Chicago Press, Chicago, 1939).

\bibitem{ShapiroTeukolsky} 
S.L. Shapiro and S.A. Teukolsky, Black Holes, White Dwarfs, and Newtron Stars, Wiley, New York, 1983.

{
\bibitem{NashedSaridakis:2020}
G.G.L. Nashed and E.N. Saridakis,
New rotating black holes in nonlinear Maxwell $f(\cal{R})$ gravity, 
Phys. Rev. D {\bf 102}, 124072 (2020).

\bibitem{Kagramanova:2006}
V. Kagramanova, J. Kunz, C. L\"ammerzahl, 
Solar system effects in Schwarzschild–de Sitter spacetime,
Phys. Lett. B {\bf 634}, 465 (2006).

\bibitem{Jetzer:2006}
Ph. Jetzer and M. Sereno,
Two-body problem with the cosmological constant and observational constraints,
Phys. Rev. D {\bf 73}, 044015 (2006).

\bibitem{Sereno:2006}
M. Sereno and Ph. Jetzer, 
Solar and stellar system tests of the cosmological constant,
Phys. Rev. D {\bf 73}, 063004 (2006).

\bibitem{Iorio:2008}
L. Iorio, Solar System Motions and the Cosmological Constant: A New Approach. Adv. Astron. {\bf 2008}, 268647 (2008). 

\bibitem{IorioSaridakis:2012}
L. Iorio, E.N. Saridakis, Solar system constraints on $f(T)$ gravity. Mon. Not. R. Astron. Soc. {\bf 427}, 1555–1561 (2012).

\bibitem{Arakida:2013}
H. Arakida, Note on the Perihelion/Periastron Advance Due to Cosmological Constant. Int. J. Theor. Phys. {\bf 52}, 1408–1414 (2013).

\bibitem{XieDeng:2013}
Y. Xie, X.M. Deng, $f(T)$ gravity: Effects on astronomical observations and Solar system experiments and upper bounds. Mon. Not. R. Astron. Soc. {\bf 433}, 3584–3589 (2013).

\bibitem{Iorio:2015}
L. Iorio, N. Radicella, M.L. Ruggiero, Constraining $f(T)$ gravity in the Solar System. J. Cosmol. Astropart. Phys. {\bf 8}, 021 (2015).

\bibitem{Iorio:2018}
L. Iorio, Perspectives on constraining a cosmological constant-type parameter with pulsar timing in the Galactic Center, Universe {\bf 4}(4), 59 (2018). 
}


\bibitem{JamesLattimer2012} 
J. M. Lattimer, The Nuclear Equation of State and Neutron Star Masses, 
Annu. Rev. Nucl. Part. Sci.  \textbf{62}, pp.485-515 (2012). 






%
%
%
%
\bibitem{JamesLattimer2014}
J. M. Lattimer, A. W. Steiner, Neutron Star Masses and Radii from Quiescent Low-Mass X-ray Binaries, The Astrophysical Journal \textbf{784}, no.2 (2014). 
\bibitem{Heinke:2006} 
C. O. Heinke, G. B. Rybicki, R. Narayan,  Jonathan E. Grindlay, 
A Hydrogen Atmosphere Spectral Model Applied to the Neutron Star X7 in the Globular Cluster 47 Tucanae, 
The Astrophysical Journal \textbf{644}, no. 2 (2006).
\bibitem{Suleimanov:2011}
V. Suleimanov, J. Poutanen, M. Revnivtsev, K. Werner, 
Neutron star stiff equation of state derived from cooling phases of the X-ray burster 4U 1724-307, 
Astrophysical Journal \textbf{742}, no.2 (2011).
\bibitem{Suleimanov:2016}
V.F. Suleimanov,  J. Poutanen, D. Klochkov, K. Werner, 
Measuring the basic parameters of neutron stars using model atmospheres. Eur. Phys. J. A \textbf{52}, 20 (2016). 
\bibitem{Miller:2017} 
J. N\"{a}ttil\"{a}, M. C. Miller, A. W. Steiner, J. J. E. Kajava, V. F. Suleimanov,  J. Poutanen,
Neutron star mass and radius measurements from atmospheric model fits to X-ray burst cooling tail spectra, 
Astron. Astrophys.  \textbf{608}, A31 (2017). 
\bibitem{Suleimanov:2017}
V. F. Suleimanov, J. J. E. Kajava, S. V. Molkov, J. N\"{a}ttil\"{a}, A. A. Lutovinov, K. Werner, J. Poutanen, Basic parameters of the helium-accreting X-ray bursting neutron star in 4U 1820-30, Monthly Notices of the Royal Astronomical Society  \textbf{472}, iss. 4, pp. 3905-3913 (2017). 
\bibitem{bookAstrNS2018}
The Physics and Astrophysics of Neutron Stars, ed L. Rezzolla et al. Springer, Cham  \textbf{457} (2018). 

\bibitem{AthanasiosKottas2013}     
B. Kiziltan, A, Kottas, M. De Yoreo, S. E. Thorsett, The Neutron Star Mass Distribution, 
The Astrophysical Journal \textbf{778}, no.1 (2013). 
\bibitem{NiranjanKumar2022} 
N. Kumar, V. V. Sokolov, 
Mass Distribution and ‘Mass Gap’ of Compact Stellar Remnants in Binary Systems,    arXiv:2204.07632. 
{
\bibitem{Doroshenko_etal:2022} 
V. Doroshenko, V. Suleimanov, G. P\"uhlhofer, et al. A strangely light neutron star within a supernova remnant. Nat. Astron. (2022). https://doi.org/10.1038/s41550-022-01800-1
}

{
\bibitem{ReggeWheeler} 
T. Regge and J. A. Wheeler, Phys. Rev. {\bf 108}, 1063 (1957).


\bibitem{Zerilli} 
F. J. Zerilli, Phys. Rev. Lett. {\bf 24}, 737 (1970).

\bibitem{KobMotSuy:I2012} 
T. Kobayashi, H. Motohashi, and T. Suyama, 
Black hole perturbation in the most general scalar-tensor theory with second-order field equations
I: The odd-parity sector, Phys. Rev. D 85, 084025 (2012), [Erratum:
Phys. Rev. D 96, 109903 (2017)].

\bibitem{KobMotSuy:II2014} 
T. Kobayashi, H. Motohashi, and T. Suyama, 
Black hole perturbation in the most general scalar-tensor theory with second-order field equations II: the even-parity sector, 
Phys. Rev. D {\bf 89}, 084042 (2014),
arXiv:1402.6740 [gr-qc].

\bibitem{OgaKobSuy:2016} 
Hiromu Ogawa, Tsutomu Kobayashi, Teruaki Suyama, 
Instability of hairy black holes in shift-symmetric Horndeski theories, 
Phys. Rev. D {\bf 93}, 064078 (2016), 
arXiv:1510.07400 [gr-qc].

\bibitem{TakSui:2017} 
Kazufumi Takahashi, Teruaki Suyama, 
Linear perturbation analysis of hairy black holes in shift-symmetric Horndeski theories: Odd-parity perturbations, 
Phys. Rev. D {\bf 95}, 024034 (2017), arXiv:1610.00432 [gr-qc].

\bibitem{TatFer:2018} 
O. J. Tattersall and P. G. Ferreira, 
Quasinormal modes of black holes in Horndeski gravity, 
Phys. Rev. D {\bf 97}, 104047 (2018), arXiv:1804.08950 [gr-qc].

\bibitem{KasKimSatTsu:2020} 
R. Kase, R. Kimura, S. Sato and S. Tsujikawa, 
Stability of relativistic stars with scalar hairs, 
Phys. Rev. D {\bf 102}, 084037 (2020),
[arXiv:2007.09864 [gr-qc]].

\bibitem{TomKob:2021} 
K. Tomikawa and T. Kobayashi, 
Perturbations and quasi-normal modes of black holes with time-dependent scalar hair in shift-symmetric scalar-tensor theories, 
Phys. Rev. D {\bf 103}, 084041 (2021), arXiv:2101.03790 [gr-qc].

\bibitem{LanHouRou:2021} 
D. Langlois, K. Noui and H. Roussille, 
Black hole perturbations in modified gravity, 
Phys. Rev. D {\bf 104}, 124044 (2021), arXiv:2103.14750 [gr-qc].

\bibitem{MinTakTsu:2022a} 
Masato Minamitsuji, Kazufumi Takahashi, Shinji Tsujikawa, 
Linear stability of black holes in shift-symmetric Horndeski theories
with a time-independent scalar field, 
Phys. Rev. D {\bf 105}, 104001 (2022),
arXiv:2201.09687 [gr-qc].

\bibitem{MinTakTsu:2022b} 
Masato Minamitsuji, Kazufumi Takahashi, Shinji Tsujikawa, 
Linear stability of black holes with static scalar hair in full Horndeski theories: generic instabilities and surviving models, 
Phys. Rev. D {\bf 106}, 044003 (2022), 
arXiv:2204.13837 [gr-qc].

\bibitem{KasTsu:2021} 
R. Kase, S. Tsujikawa, 
Instability of compact stars with a nonminimal scalar-derivative coupling, 
JCAP {\bf 01}, 008 (2021), arXiv:2008.13350 [gr-qc].

\bibitem{KasTsu:2022} 
R. Kase, S. Tsujikawa, 
Relativistic star perturbations in Horndeski theories with a gauge-ready formulation, 
Phys. Rev. D {\bf 105}, 024059 (2022), 
arXiv:2110.12728 [gr-qc].

\bibitem{KhoNouTroWong:2022} 
Justin Khoury, Toshifumi Noumi, Mark Trodden, Sam S. C. Wong, 
Stability of Hairy Black Holes in Shift-Symmetric Scalar-Tensor Theories
via the Effective Field Theory Approach, 4 Aug 2022, arXiv:2208.02823
[hep-th].

\bibitem{MinTsu:2022}
M. Minamitsuji, S. Tsujikawa, 
Stability of neutron stars in Horndeski theories with Gauss-Bonnet couplings,
arXiv:2207.04461.

\bibitem{inprep}
P. E. Kashargin, S. V. Sushkov,
Stability of anti-de Sitter neutron stars in the theory of gravity with nonminimal derivative coupling,
in preparation.
}

\end{thebibliography}
\end{document}